\documentclass[journal]{IEEEtran}

\usepackage{cite}

\usepackage{soul}
\usepackage{graphicx}
\usepackage{epstopdf}
\usepackage{pdflscape}
\DeclareGraphicsExtensions{.eps}
\usepackage{array}
\usepackage{hhline}
\usepackage{multirow}
\usepackage{booktabs}
\usepackage{stfloats}
\usepackage{algorithm}
\usepackage{algorithmic}

\usepackage[thmmarks]{ntheorem}

\makeatletter
\newtheoremstyle{MyNonumberplain}%
  {\item[\theorem@headerfont\hskip\labelsep ##1\theorem@separator]}%
  {\item[\theorem@headerfont\hskip\labelsep ##3\theorem@separator]}
\makeatother
\theorembodyfont{\upshape}

\usepackage[cmex10]{amsmath}
\usepackage{amsfonts,amsmath,amssymb,epsfig,epstopdf, color}
\makeatletter
\renewcommand{\maketag@@@}[1]{\hbox{\m@th\normalsize\normalfont#1}}%
\makeatother
\DeclareMathOperator{\diag}{diag}

\newcommand{\norm}[1]{\left\lVert#1\right\rVert}

\usepackage[tight,footnotesize]{subfigure}

\begin{document}
\title{Reduced Switching Connectivity for \\ Large Scale Antenna Selection}
\author{\IEEEauthorblockN{Adrian Garcia-Rodriguez,~\IEEEmembership{Member,~IEEE}, Christos Masouros,~\IEEEmembership{Senior Member,~IEEE}, and Pawel Rulikowski}
\thanks{This work was supported by the Royal Academy of Engineering, UK and the EPSRC under grant EP/M014150/1.}
\thanks{A.\ Garcia-Rodriguez and P.\ Rulikowski are with Nokia Bell Labs, Dublin 15, Ireland. E-mail: \{adrian.garcia$\_$rodriguez, pawel.rulikowski\}@nokia-bell-labs.com.}
\thanks{C.\ Masouros is with the Department of Electronic \& Electrical Engineering, University College London, London WC1E 7JE, U.K. E-mail: c.masouros@ucl.ac.uk. A.\ Garcia-Rodriguez was also at University College London.}
\thanks{“\copyright 2017 IEEE. Personal use of this material is permitted. Permission from IEEE
must be obtained for all other uses, including reprinting/republishing this material for
advertising or promotional purposes, collecting new collected works for resale or
redistribution to servers or lists, or reuse of any copyrighted component of this work in other
works.}
\thanks{Digital Object Identifier 10.1109/TCOMM.2017.2669030.}
\thanks{The published version of the article can be found at: http://ieeexplore.ieee.org/document/7855725/}
}
\maketitle

\begin{abstract}
In this paper, we explore reduced-connectivity radio frequency (RF) switching networks for reducing the analog hardware complexity and switching power losses in antenna selection (AS) systems. In particular, we analyze different hardware architectures for implementing the RF switching matrices required in AS designs with a reduced number of RF chains. We explicitly show that fully-flexible switching matrices, which facilitate the selection of any possible subset of antennas and attain the maximum theoretical sum rates of AS, present numerous drawbacks such as the introduction of significant insertion losses, particularly pronounced in massive multiple-input multiple-output (MIMO) systems. Since these disadvantages make fully-flexible switching suboptimal in the energy efficiency sense, we further consider partially-connected switching networks as an alternative switching architecture with reduced hardware complexity, which we characterize in this work. In this context, we also analyze the impact of reduced switching connectivity on the analog hardware and digital signal processing of AS schemes that rely on received signal power information. Overall, the analytical and simulation results shown in this paper demonstrate that partially-connected switching maximizes the energy efficiency of massive MIMO systems for a reduced number of RF chains, while fully-flexible switching offers sub-optimal energy efficiency benefits due to its significant switching power losses.
\end{abstract}

\IEEEpeerreviewmaketitle

\section{Introduction}

\IEEEPARstart{T}{HE} constant growth in the number of mobile devices as well as the development of data-hungry applications have driven the design of novel wireless communications solutions. In this context, systems incorporating a large number of antennas at the base stations (BSs) and occupying larger bandwidths are leading candidates for 5G \cite{6736746,6824752}. In particular, massive MIMO for conventional microwave frequencies and millimeter waves implement a large number of antennas for achieving unprecedented spatial resolutions and for compensating for the larger signal attenuations introduced at high frequencies \cite{6375940, 5783993}. Simultaneously, these solutions motivate the design of novel digital and analog signal processing strategies, since the economic cost, power consumption, and hardware and digital signal processing complexities of systems with a dedicated RF chain per antenna might become burdensome \cite{7402270, 5783993,5595728}. In order to overcome the above challenges, a number of solutions such as hybrid analog-digital precoding schemes \cite{5783993} and spatial modulation  \cite{6678765, 7109850} have been proposed. In this context, AS has also been posed as a feasible alternative for reducing the complexity in both small and large scale MIMO systems operating below 6 GHz \cite{1227919, 1284943, 1341263}, and comprises the \mbox{focus of this work.}

The characterization and development of AS systems have constituted the focus of several works, which study different features such as the acquisition of channel state information (CSI) \cite{4633647, 6493985}, their energy efficiency improvements \cite{7100915,6725592}, or their practical implementation aspects \cite{4149881, 4564614}. The implementation of AS as a means of reducing the excessive number of RF chains in massive MIMO has also attracted considerable attention \cite{7247768, 6725592, 6824762, 7390948, 7172496, 7417765}. Particularly relevant in this context are the results presented in \cite{7172496, 7417765}, which illustrate that a large portion of the full-RF massive MIMO rates can be attained via AS in realistic propagation scenarios. However, none of the above-mentioned works presents a comprehensive analysis of a critical aspect: the design of the RF switching matrix that implements the AS.

The RF switching matrix represents the hardware components required in AS for interconnecting the RF chains with their selected antennas in implementations with a reduced number of RF chains \cite{1284943, 4149881}. The design of RF switching matrices has been mostly studied in the area of microwave circuit implementation. Illustrative examples in this line include \cite{1381671, 4956996, 6546638, 5944901, nazemzadeh2008guide, teledyineSwitchingMatrices}, where the focus is on fully-flexible switching matrices that facilitate the inter-connection of any input port to all output ports. The collosal growth of antenna numbers in massive MIMO also increases the complexity of RF switching, hence making these components a significant performance factor. In spite of acknowledging the crucial importance of this component \cite{1284943}, the impact of the switching network on the ergodic sum rates and energy efficiency of AS has been commonly ignored in the related literature. Only recently, the large insertion losses and complexity of switching matrices in massive MIMO have been considered in \cite{7417765} and \cite{7370753}. Specifically, \cite{7370753} considers a switching network that only connects each RF chain to predefined subsets of antennas for reducing the complexity of switching between the large number of antennas required at millimeter wave frequencies. Similarly, \cite{7417765} proposes to implement the switching matrix via binary switches to alleviate the insertion losses. However, this comes at the expense of reducing the input-output connectivity of the resultant switching architecture, which is partially-connected. Still, these works do not perform a thorough and detailed analysis of the hardware implications behind their designs and only focus on specific implementations such as binary switching matrices.

Considering the above, in this contribution we generalize the above-mentioned works to arbitrary switching architectures and concentrate on providing a detailed analysis of switching networks to characterize their influence in AS systems. In particular, we present a number of specific hardware implementations of switching matrices that are optimized under different criteria such as the number of internal connections or the signal power losses. In this context, we accurately characterize their insertion losses, which are shown to be critical for massive MIMO due to the large number of inputs (RF chains) and outputs (antennas) required. In contrast with \cite{1381671, 4956996, 6546638, 5944901, nazemzadeh2008guide, teledyineSwitchingMatrices}, we also consider architectures with limited connectivity as a means for reducing the complexity of the fully-flexible switching networks conventionally considered. In this line, we also determine the ergodic sum capacity loss introduced due to the limited connectivity for received signal power-based AS systems, since power-based AS can offer a superior performance when practical CSI acquisition procedures are considered. Altogether, the results and generalized designs considered in this work provide a comprehensive view of the impact of RF switching matrices on the maximum sum achievable rates, hardware architecture and energy efficiency of AS systems.

The rest of the paper is structured as follows. Sec.\ \ref{sec:globalSystemModel} reviews the fundamentals of AS systems, including their CSI acquisition procedure. Sec.\ \ref{sec:FFSwitching} and Sec.\ \ref{sec:PCSwitching} characterize the operation of fully-flexible and partially-connected switching matrices, respectively. Sec.\ \ref{sec:performanceAnalysis} provides theoretical approximations of the ergodic sum capacity achieved by partially-connected AS systems. Sec.\ \ref{sec:EEAnalysis} introduces the energy efficiency metric employed in Sec.\ \ref{sec:simulationResults}, where numerical results are presented. The conclusions of this paper are finally drawn in Sec.\ \ref{sec:conclusion}.

\section{Preliminaries}
\label{sec:globalSystemModel}

\subsection{Downlink System Model}

Let us consider a time-division duplex (TDD) multi-user MIMO (MU-MIMO) system comprised of a BS with $N$ transmit antennas and $K$ single-antenna mobile stations (MSs). The BS incorporates $M \in \left\{ K, K+1, \ldots, N \right\}$ RF chains to convey $K$ independent data symbols to the MSs via AS. The baseband representation of the composite signal $\mathbf{y} \in \mathbb{C}^{K \times 1}$ received by the MSs can be expressed as
\begin{equation}
\mathbf{y}  = \sqrt{\rho} \mathbf{H}_{\left[ \mathcal{M} \right]} \mathbf{x} + \mathbf{w},
\label{eq:systemModel}
\end{equation}
where $\mathbf{H} \in \mathbb{C}^{K \times N}$ denotes the channel response between the BS antennas and the users, $\mathbf{x} \in \mathbb{C}^{M \times 1}$ is the transmit signal and $\mathbf{w}\in\mathbb{C}^{K \times1}\sim\mathcal{C}\mathcal{N}(\mathbf{0}, \textbf{I}_{K})$. In the previous expressions, $\sim$ means ``distributed as'', $\textbf{I}_{K}$ is the $K \times K$ identity matrix and $\mathcal{C}\mathcal{N}( \mathbf{a}, \mathbf{A} )$ represents a complex normally distributed vector with means $\mathbf{a}$ and covariance matrix $\mathbf{A}$. Moreover, $\mathbf{H}_{\left[ \mathcal{M} \right]} \in \mathbb{C}^{K \times M}$ is a submatrix of $\mathbf{H}$ built by selecting the columns specified by the set $\mathcal{M} \subseteq \left\{ 1, \ldots , N \right\}$ with cardinality $\vert \mathcal{M} \vert = M$. Here, $\mathcal{A} \subseteq \mathcal{B}$ indicates that $\mathcal{A}$ is a subset of $\mathcal{B}$. We let $\mathbb{E} \left[ \mathbf{x}^{\rm H} \mathbf{x} \right] = K$, where $\mathbb{E} \left[ \cdot \right]$ denotes statistical expectation. Based on the above, $\rho$ represents the average transmission power per MS.

%\subsection{Antenna Selection Benchmarks}
%\label{sec:antennaSelectionMIMO}

\begin{figure}[!t]
	\centering
		\includegraphics[width=0.45\textwidth]{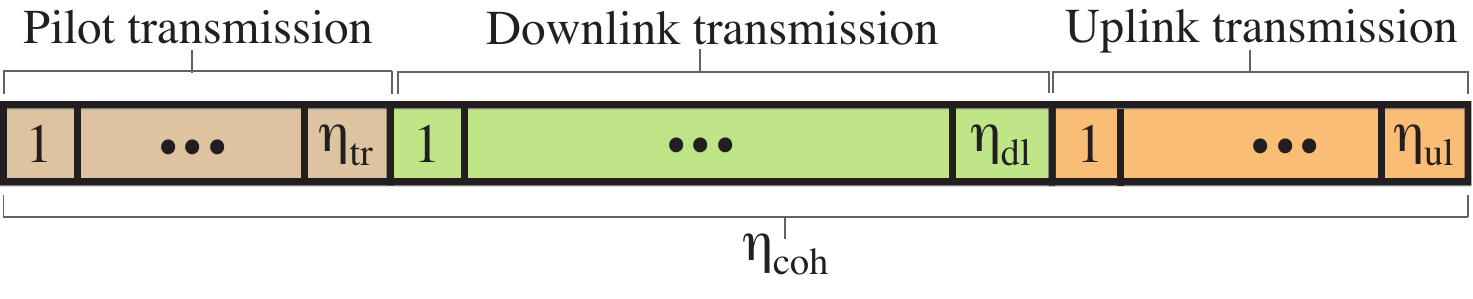}
		\caption{Operation of a TDD communication system.}
\label{fig:TDDOperation}
\end{figure}

The sum capacity $C$ of the MU-MIMO downlink system described in \eqref{eq:systemModel} is given by \cite{1203154,4350224}
\begin{equation}
C = \underset{\mathbf{P}}{\max} \mbox{ } \left( \frac{\eta_{\rm dl}}{\eta_{\rm coh}} \right)  \log_{2} \det \left( \mathbf{I}_{K} + \rho \mathbf{P}  \mathbf{H}_{\left[ \mathcal{M} \right]} \left( \mathbf{H}_{\left[ \mathcal{M} \right]}\right)^{\rm H} \right),
\label{eq:capacityDL}
\end{equation}
where $\left( \cdot \right)^{\rm H}$ denotes the Hermitian transpose of a matrix, whereas $\eta_{\rm coh}$ and $\eta_{\rm dl}$ denote the total and downlink time-frequency symbols per channel coherence interval, respectively, as shown in \figurename~\ref{fig:TDDOperation}. Moreover, $\mathbf{P} \in \mathbb{R}^{K \times K}$ is a diagonal user power allocation matrix satisfying $\sum_{i = 1}^{K} \mathbf{P}_{i,i} = K$, where $\mathbf{A}_{i,j}$ denotes the $i,j$-th entry of $\mathbf{A}$. 

% Moreover, inspired by \cite{7172496}, in this work we also consider the practical linear zero-forcing (ZF) precoder, whose achievable sum rates can be obtained as the resultant value of the objective function in the optimization problem
%\begin{align}
%\mathcal{P}_{2}:  \mbox{ } & \underset{\mathbf{D}}{\text{maximize }} \left( \frac{\eta_{\rm dl}}{\eta_{\rm coh}} \right)  \sum_{i=1}^{K} \log_{2} \left( 1 + \rho \mathbf{D}_{i,i}  \right) \label{eq:ZFRates} \\
%& \text{subject to } \sum_{i=1}^{K}  \mathbf{D}_{i,i} \left( \left( \mathbf{H} \mathbf{\widetilde{S}}^{\star} \mathbf{H}^{\rm H}  \right)^{-1} \right)_{i,i}  = K, \nonumber
%\end{align}
%where $\left( \cdot \right)^{-1}$ denotes the inverse matrix and $\mathbf{D} \in \mathbb{R}^{K \times K}$ is the diagonal user power allocation matrix for ZF. The solution to $\mathcal{P}_{2}$ can also be obtained via conventional water-filling \cite{cover2012elements}. Moreover, note that $\mathcal{P}_{2}$ employs the possibly suboptimal $\mathbf{\widetilde{S}}^{\star}$ obtained from $\mathcal{P}_{1}$, which is in line with \cite{7172496, 7417765,917094}.

\subsection{Channel State Information Acquisition for AS Systems}
\label{sec:CSIAcquisition}

The sum capacity $C$ in \eqref{eq:capacityDL} can only be achieved provided that full knowledge of the channel response $\mathbf{H}$ is available at the BS \cite{7172496, 1687757}. However, CSI acquisition poses a major challenge in AS systems with reduced RF chains, which constitutes the focus of this section. In order to characterize this issue we adopt a standard block fading model in the following, i.e., the channel is considered to remain approximately constant throughout a block comprised of $\eta_{\rm coh}$ time-frequency symbols, and to vary independently between blocks \cite{7031971}.

\subsubsection{Instantaneous CSI Acquisition for reduced-RF systems ($M<N$)}

The AS systems considered in this paper only implement $M < N$ RF chains. This constraint entails that only $M$ signals from the antenna ports can be processed simultaneously. In other words, a multiplexed training stage is required to estimate the channels of all $N$ antennas \cite{4633647}. Specifically, the minimum number of training symbols per coherence interval required to estimate the complete channel is given by
\begin{equation}
\eta_{\rm{tr}} = K \times \left\lceil \frac{N}{M} \right\rceil, 
\label{eq:channelAcquisitionRequired}
\end{equation}
where $\lceil \cdot \rceil$ rounds to the highest closer integer.  The resultant number of time-frequency symbols dedicated to downlink data transmission per coherence block is given by\footnote{For simplicity, we ignore the guard intervals between the uplink, downlink and pilot transmission stages found in realistic TDD systems.}
\begin{equation}
\eta_{\rm{dl}} = \eta_{\rm{coh}} - \eta_{\rm{ul}} - \eta_{\rm{tr}} =  \eta_{\rm{coh}} - \eta_{\rm{ul}} -  \eta_{\rm{tr}},
\label{eq:channelAcquisitionRequiredFullCSI}
\end{equation}
where $\eta_{\rm{ul}}$ refers to the number of uplink data symbols in each coherence block, as illustrated in \figurename~\ref{fig:TDDOperation}.

\subsubsection{Power-Based AS for reduced-RF systems ($M<N$)}

An elegant solution to the CSI acquisition problem in AS systems consists in adopting a selection decision hinging on the norm of the channel entries. With this purpose, let $\overline{\mathbf{h}} = \left[ \norm{\mathbf{h}_{1}}^{2}, \cdots, \norm{\mathbf{h}_{N}}^{2} \right] \in \mathbb{R}^{1 \times N}$ be a vector comprising the channel power measured per antenna element, where $\mathbf{h}_{i}$ denotes the $i$-th column of $\mathbf{H}$. Let $\mathbf{S} \in \mathbb{B}^{N \times N}$ be a diagonal binary matrix that indicates whether the $i$-th antenna element is selected, i.e.
\begin{equation}
\mathbf{S}_{i,i} = \begin{cases}
1, \mbox{ } \text{ if } i \in \mathcal{M}, \\
0, \mbox{ } \text{ otherwise}.
\end{cases}
\label{eq:selectionMatrixDefinition}
\end{equation}
For the case of power-based AS, $\mathbf{S}$ can be obtained by solving
\begin{align}
\mathcal{P}_{1}: & \underset{\mathbf{S}}{\text{ maximize }} \overline{\mathbf{h}} \mathbf{S} \label{eq:originalPBASOptimization} \\
& \text{ subject to } \sum_{i = 1}^{N} \mathbf{S}_{i,i} = M, \nonumber \\
& \kern 4.6em  \mathbf{S}_{i,i} \in \left\{ 0, 1 \right\}, \mbox{ } \forall \mbox{ } i \in \left\{ 1, \ldots , N \right\}. \nonumber
\end{align}
This strategy is commonly referred to as power-based or norm-based AS. The solution to the power-based AS of $\mathcal{P}_{1}$ is straightforward for the case of fully-flexible architectures, i.e.\ \cite{1341263}
\begin{equation}
\widetilde{\mathbf{S}}^{\star} = \max _{M} \norm{\mathbf{h}_{i}}^{2},
\label{eq:PBSelectionWithFC}
\end{equation}
where $\displaystyle\max_{M} \left( \cdot \right)$ selects the largest $M$ entries. Power-based AS generally provides a sub-optimal decision for maximizing the sum capacity of MU-MIMO systems, since the latter does not only depend on the received signal power but also on the orthogonality between the channels of different users \cite{1284943}. However, power-based AS is also capable of reducing the amount of time resources spent for CSI acquisition in systems where RF power meters instead of full RF chains are attached to each antenna port \cite{4633647}. This is because a) the channel power information can be acquired from the prior uplink stage and b) this information can be subsequently employed for AS as per \eqref{eq:originalPBASOptimization}. The above entails that power-based AS only requires a minimum of $\eta_{\rm tr} = K$ training pilots \cite{5595728}.

\subsection{Sources of Losses in the RF Switching Matrices of AS}

The design of the RF switching matrix in AS plays a fundamental role in the overall system performance \cite{1284943}. Among the multiple technical aspects that should be considered from a system-level design perspective, the most relevant ones are:
\begin{itemize}
\item \emph{Insertion losses.} Insertion losses account for the signal power losses introduced by RF switching matrices when the input power is not fully delivered to the output ports \cite{pozar2009microwave}. These losses generally grow with the number of input and output ports \cite{nazemzadeh2008guide,teledyineSwitchingMatrices}. Therefore, this is a critical parameter in massive MIMO, due to the large number of RF chains and antennas implemented \cite{6375940, 7417765}.
\item \emph{Coupling between ports.} The coupling of the switching matrix determines the fraction of the signals that appear at a specific port, but were intended for other ports. This parameter depends on the network of connections inside the switching matrix and the power leakage of the internal switching devices (e.g., FET transistors or electromechanical switches) \cite{niknejad2007electromagnetics}.
\item \emph{Transfer function balance.} The transfer function of each input-output combination between RF chain and antenna element should be ideally identical to ensure that the baseband model \eqref{eq:systemModel} accurately characterizes the system's operation \cite{1284943}.
\end{itemize}

From the above three sources of losses, here we focus on the insertion losses. This is because a) transfer function imbalances can be compensated by means of an initial system calibration \cite{1284943} and b) coupling between internal switching ports can be in the order of $-20$ to $-30$ dB, hence effectively making unintentional power transfers between nearby ports negligible \cite{peregrineSwitches}. Instead, the insertion losses introduced by switching matrices with a large number of input and output ports can be in the order of $2$-$3$ dB \cite{teledyineSwitchingMatrices}, hence dominating the overall performance loss. For this reason, Sec.\ \ref{sec:FFSwitching} explores a number of implementations of conventional fully-flexible switching matrices, with emphasis on the number of switching components required and the associated insertion losses.

\section{Fully-Flexible Switching for Antenna Selection}
\label{sec:FFSwitching}

\begin{figure}[!t]
	\centering
		\includegraphics[width=0.49\textwidth]{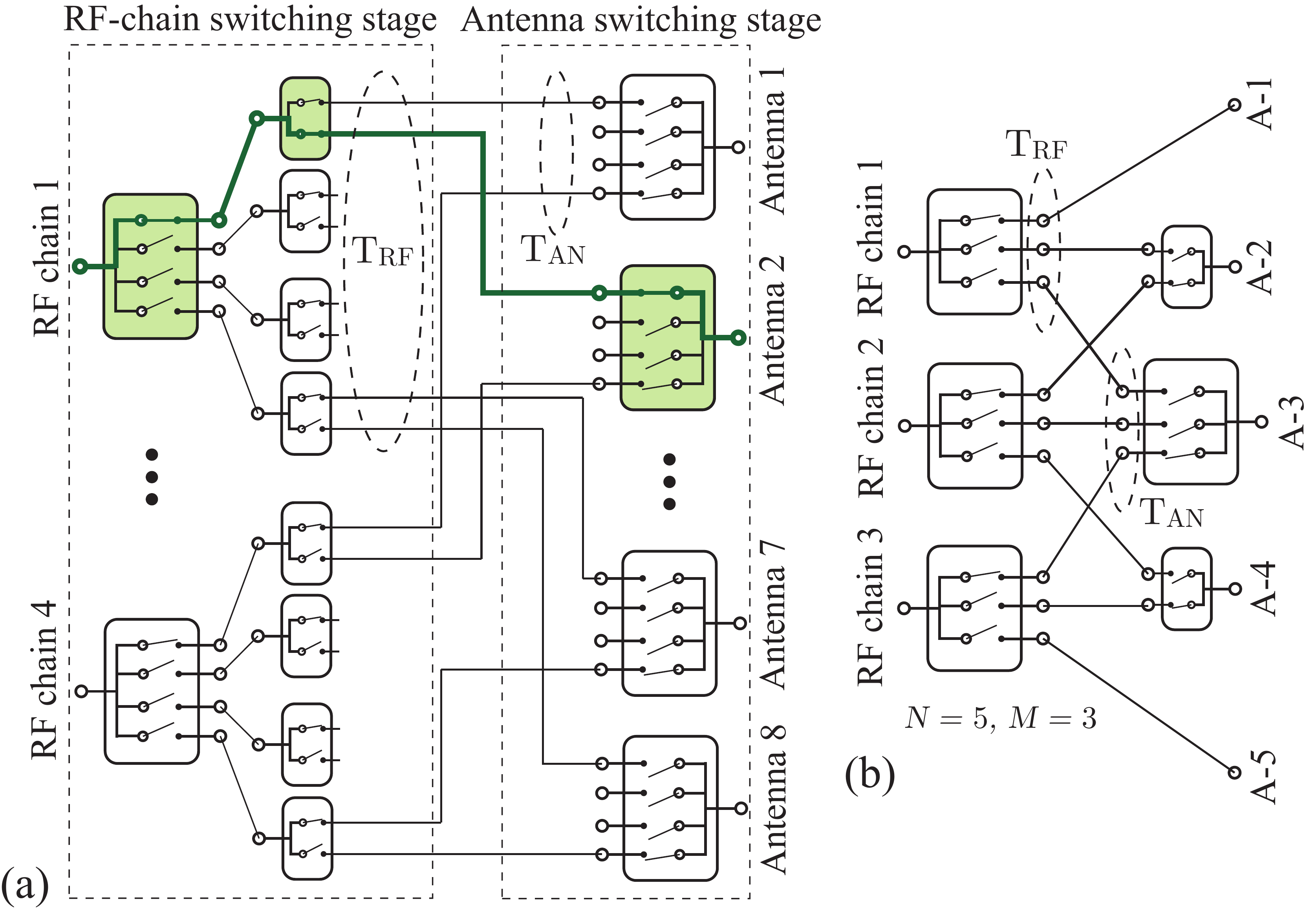}
		\caption{Block diagrams of fully-flexible architectures: (a) a conventional $4 \times 8$ switching matrix with each input port connected to every output port (full flexibility with full connectivity) and (b) a switching matrix minimizing the number of internal connections (full flexibility with minimum connectivity).}
\label{fig:Conventional4x8}
\end{figure}

RF switching matrices in conventional AS have two essential requirements: a) connecting each RF chain to the antenna ports (full flexibility) and b) allowing bidirectional switching for uplink-downlink operation \cite{nazemzadeh2008guide,1284943}. An illustrative example of a switching matrix is shown in \figurename~\ref{fig:Conventional4x8}(a), where the block diagram of a $4 \times 8$ architecture can be observed\footnote{Although not analyzed in this paper for reasons of space, alternative switching architectures for large scale antenna systems could also be implemented (see, e.g., \cite{4014841}).}. This figure shows that a large switching matrix is comprised of two main switching stages (represented by the dashed boxes in the figure) with multiple switches of smaller size: one stage at the RF-chain ports, referred to as RF-chain switching stage in the sequel, and a subsequent stage at the antenna ports, referred to as antenna switching stage.

Each of the above two switching stages is comprised by a number of switches with a smaller number of ports, which will be referred to as \emph{basic switches} in the following. Indeed, the RF-chain switching stage of \figurename~\ref{fig:Conventional4x8}(a) illustrates that several of these basic switches can be concatenated to produce the desired number of ports \cite{teledyineSwitchingMatrices}. The basic switches considered in this paper conventionally follow the nomenclature SPXT (single-pole X-throw) which refers to the number of separate ports with independent signals that the basic switch can control (poles) and the number of different signal paths that the switch allows for each pole (throws) \cite{peregrineSwitches,skyworksSwitches}. For instance, a SP3T switch is capable of routing one signal from or towards three different ports (throws). Without loss of generality, we consider the SP2T-SP4T basic switches with their insertion loss detailed in Table \ref{tab:basicSwitches} for the illustrative architectures \mbox{shown in this paper \cite{peregrineSwitches}.}

\small
\begin{table}[!t] 
\begin{center}
{\renewcommand{\arraystretch}{1.4}
\caption{Basic SPXT switches}
\begin{tabular}{| p{1.5cm} | p{1.5cm}| p{2.3cm}|} 
 \hline
{\bf \emph{Switch type}} & {\bf \emph{Model}} \cite{peregrineSwitches} & {\bf \emph{Insertion loss}}
\\ \hline \hline
SP$2$T & PE42422  & $L_{2} = 0.25$ dB
\\ \hline
SP$3$T & PE42430 & $L_{3} = 0.45$ dB
\\ \hline
SP$4$T & PE42440 & $L_{4} = 0.45$ dB
\\ \hline
\end{tabular}
\vspace{-0.4cm}
\label{tab:basicSwitches}}
\end{center}
\end{table}\normalsize
 
Conventional AS systems consider fully-flexible switching matrices that allow any possible combination of $M$ antennas to be simultaneously selected. However, a number of implementations for the design of switching matrices with varying complexity and insertion loss can be implemented. Importantly, different architectures may result in RF-chain and antenna switching stages with different number of ports (throws). An accurate characterization of the maximum number of throws per switching stage is crucial, since they determine overall insertion loss of the critical signal path, i.e. the signal path with largest power losses.

With the above purpose, let us define $T_{\text{RF}}$ and $T_{\text{AN}}$ as the maximum number of throws per RF chain ($T_{\text{RF}}$) in the RF-chain switching stage or per antenna ($T_{\text{AN}}$) in the antenna switching stage, as represented in \figurename~\ref{fig:Conventional4x8}(a). Moreover, let $\mathcal{T}$ be the set with elements in decreasing order given by the number of throws in the basic switches, i.e., $\mathcal{T}  = \left\{ 4, 3, 2 \right\}$ for the basic switches considered in Table \ref{tab:basicSwitches}. $\mathcal{T}_{j}$ corresponds to the $j$-th entry in $\mathcal{T}$ and the cardinality of $\mathcal{T}$ is $\vert \mathcal{T} \vert \triangleq N_{\rm s}$. Intuitively, $N_{\rm s}$ refers to the number of different basic switches considered, i.e. $N_{\rm s} = 3$ for the basic switches of Table \ref{tab:basicSwitches}. The total insertion loss measured in dB of the critical signal path $L$ for a given switching architecture can be computed as
\begin{align}
L & = \sum_{j = 1}^{N_{\rm s}} \left( S_{\mathcal{T}_{j}}^{\text{RF}} + S_{\mathcal{T}_{j}}^{\text{AN}} \right) \times L_{\mathcal{T}_{j}} = \sum_{j = 1}^{N_{\rm s}}  S_{\mathcal{T}_{j}} \times L_{\mathcal{T}_{j}},
\label{eq:totalIL}
\end{align}
where $S_{\mathcal{T}_{j}}^{\text{RF}}$ and $S_{\mathcal{T}_{j}}^{\text{AN}}$ represent the number of consecutive basic switches with $\mathcal{T}_{j}$ throws that the signals cross in the RF-chain and in the antenna switching stages, respectively, and $L_{\mathcal{T}_{j}}$ denotes the insertion loss in dB introduced by a basic switch with $\mathcal{T}_{j}$ throws. For instance, we have $L_{2} = 0.25$ dB, $L_{3} = 0.45$ dB and $L_{4} = 0.45$ dB for the basic switches considered in Table \ref{tab:basicSwitches}. Moreover, $S_{\mathcal{T}_{j}} \triangleq S_{\mathcal{T}_{j}}^{\text{RF}} + S_{\mathcal{T}_{j}}^{\text{AN}}$ refers to the total number of switches with $\mathcal{T}_{j}$ throws in the critical signal path of the overall switching matrix.

The number of basic switches with $\mathcal{T}_{j}$ throws crossed by the transmit signals in the RF-chain and antenna switching stages can be iteratively computed as\footnote{For simplicity, it has been considered that $T_{\text{RF}}$ and $T_{\text{AN}}$ can be factorized into the integers contained in $\mathcal{T}$. Otherwise, $T_{\text{RF}}$ and $T_{\text{AN}}$ represent the closest greater integer that can be factorized.}
\begin{equation}
S_{\mathcal{T}_{j}}^{\left\{ \text{RF, AN} \right\}} = \text{fact} \left( Q_{\mathcal{T}_{j}}^{\left\{\text{RF,AN} \right\}} , \mathcal{T}_{j} \right), j \in \left\{ 1, \ldots, N_{\rm s} \right\},
\label{eq:switchesNumber}
\end{equation}
where $\text{fact} \left( a , b \right)$ denotes the number of times $b$ appears in the integer factorization of $a$ and $Q_{\mathcal{T}_{j}}^{\left\{ \text{RF,AN} \right\}}$ is given by
\begin{equation}
\label{eq:QDefinitionAS}
Q_{\mathcal{T}_{j}}^{\left\{ \text{RF,AN} \right\}} = \begin{cases}
T_{\left\{ \text{RF,AN} \right\}}, \mbox{ } \text{ if } j = 1, \\
Q_{\mathcal{T}_{j-1}}^{\left\{ \text{RF,AN} \right\}} \Bigg/ \max \left( \mathcal{T}_{j-1}^{\left( S_{\mathcal{T}_{j-1}}^{\left\{ \rm \text{RF,AN} \right\}} \right)} , 1 \right), \text{ otherwise}.
\end{cases}
\end{equation}
In plain words, $Q_{\mathcal{T}_{j}}^{\left\{ \text{RF, AN} \right\}}$ indicates the maximum number of throws per RF-chain or per antenna that should be implemented by concatenating basic switches with $\mathcal{T}_{j}$ throws. Moreover, $\mathcal{T}_{j}^{S_{\mathcal{T}_{j}}}$ represents the number of throws obtained by $S_{\mathcal{T}_{j}}$ subsequent switching stages comprised of basic switches with $\mathcal{T}_{j}$ throws, as described in the illustrative example of Table \ref{tab:exampleArchitecture3} for the $N=128$, $M=76$ case.

\begin{table*}[!t] 
\begin{center}
{\renewcommand{\arraystretch}{1.4}
\caption{Insertion losses introduced by different switching architectures in an AS system with $N = 128$ and $M = 76$}
\begin{tabular}{| p{4.75cm} | p{2.85cm} | p{2.8cm}| p{2.7cm}| p{2.7cm}|} 
 \hline
{\bf \emph{ Parameter }} & { \emph{Full flexibility with full connectivity}} & { \emph{Full flexibility with minimum connectivity}} & { \emph{Full flexibility with minimum insertion losses}}  &  { \emph{Partial connectivity}} 
\\ \hline \hline
$T_{\text{RF}}$ & $128$ & $54$ & $64$ & $2$
\\ \hline
$T_{\text{AN}}$ & $81$ & $54$ & $64$ & $2$
\\ \hline
$Q_{\left\{ 4, 3, 2 \right\}}^{ \text{RF} }$ as per (\ref{eq:QDefinitionAS}) & $Q_{\left\{ 4,3,2 \right\}}^{ \text{RF} } = \left\{ 128,  2 , 2  \right\}$ & $Q_{\left\{ 4,3,2 \right\}}^{ \text{RF} } = \left\{ 54 , 54 , 2 \right\}$ & $Q_{\left\{ 4,3,2 \right\}}^{ \text{RF} } = \left\{ 64 , 0 , 0 \right\}$ & $Q_{\left\{ 4,3,2 \right\}}^{ \text{RF} } = \left\{ 2, 2, 2 \right\}$
\\ \hline
$\{$SP4T, SP3T, SP2T$\}$ basic switches required per RF chain $\left(S_{\left\{ 4, 3, 2 \right\}}^{\text{RF}} \right)$ & $S_{\left\{ 4,3,2 \right\}}^{\text{RF}} = \left\{ 3 , 0 , 1 \right\}$ & $S_{\left\{ 4,3,2 \right\}}^{\text{RF}} = \left\{ 0 , 3, 1 \right\}$ & $S_{\left\{ 4,3,2 \right\}}^{\text{RF}} = \left\{ 3 , 0, 0 \right\}$ & $S_{\left\{ 4,3,2 \right\}}^{\text{RF}} = \left\{ 0, 0, 1 \right\}$
\\ \hline
$Q_{\left\{ 4,3,2 \right\}}^{ \text{AN} }$ as per (\ref{eq:QDefinitionAS}) & $Q_{\left\{ 4,3,2 \right\}}^{ \text{AN} } = \left\{ 81 , 81 , 0 \right\}$ & $Q_{\left\{ 4,3,2 \right\}}^{ \text{AN} } = \left\{ 54 , 54 , 2 \right\}$ & $Q_{\left\{ 4,3,2 \right\}}^{ \text{AN} } = \left\{ 64, 0, 0 \right\}$ & $Q_{\left\{ 4,3,2 \right\}}^{ \text{AN} } = \left\{ 2, 2, 2 \right\}$
\\ \hline
$\{$SP4T, SP3T, SP2T$\}$ basic switches required per antenna $\left(S_{\left\{ 4,3,2 \right\}}^{\text{AN}} \right)$ & $S_{\left\{ 4,3,2 \right\}}^{\text{AN}}  = \left\{ 0, 4, 0 \right\}$ & $S_{\left\{ 4,3,2 \right\}}^{\text{AN}}  = \left\{ 0 , 3, 1 \right\}$ & $S_{\left\{ 4,3,2 \right\}}^{\text{AN}}  = \left\{ 3, 0, 0 \right\}$ & $S_{\left\{ 4,3,2 \right\}}^{\text{AN}}  = \left\{ 0, 0, 1 \right\}$
\\ \hline
Total number of $\{$SP4T, SP3T, SP2T$\}$ basic switches $\left( S_{\left\{ 4,3,2 \right\}} \right)$ & $S_{\left\{ 4,3,2 \right\}}  = \left\{ 3, 4, 1 \right\}$ & $S_{\left\{ 4,3,2 \right\}}  = \left\{ 0, 6, 2 \right\}$ & $S_{\left\{ 4,3,2 \right\}}  = \left\{ 6, 0, 0 \right\}$ & $S_{\left\{ 4,3,2 \right\}}  = \left\{ 0, 0, 2 \right\}$
\\ \hline
Total insertion loss in dB ($L$) as per (\ref{eq:totalIL}), where $L_{\mathcal{T}_{j}}$ are given by Table \ref{tab:basicSwitches} & $L = 3 \times 0.45 + 4 \times 0.45 + 0.25 = 3.4$ dB & $L = 6 \times 0.45 + 2 \times 0.25 = 3.2$ dB & $L = 6 \times 0.45 = 2.7$ dB & $L = 2 \times 0.25 = 0.5$ dB
\\ \hline
\end{tabular}
\vspace{-0.4cm}
\label{tab:exampleArchitecture3}}
\end{center}
\end{table*}

Considering the above, in this work we explore three implementations for the design of fully-flexible switching matrices:
\begin{enumerate}
\item \emph{Architecture $1$. Conventional fully-flexible architecture with full connectivity.} This architecture is illustrated in \figurename~\ref{fig:Conventional4x8}(a), where it can be seen that each RF chain is connected to every antenna port. In this particular case
\begin{equation}
T_{\text{RF}} = N, \text{ and } T_{\text{AN}} = M.
\label{eq:constraintsFCA1}
\end{equation}

\item \emph{Architecture $2$. Fully-flexible architecture with minimum connectivity.} This architecture minimizes the maximum number of ports at the RF-chain and antenna switching stages. The block diagram of this architecture is shown in \figurename~\ref{fig:Conventional4x8}(b) for an illustrative $3 \times 5$ RF switching matrix. Here, it can be seen that there are additional constraints regarding the connectivity of each antenna. For instance, \figurename~\ref{fig:Conventional4x8}(b) shows that the first RF chain does not connect to antenna ports $A$-$4$ and $A$-$5$. In spite of this, a full flexibility is guaranteed provided that
\begin{equation}
T_{\text{RF}} = N-M+1, \text{ and } T_{\text{AN}} = \min \left( M, T_{\text{RF}} \right).
\label{eq:constraintsFCA2}
\end{equation}
Overall, the fully-flexible with minimum connectivity architecture selects the basic switches such that $T_{\text{RF}}$ and $T_{\text{AN}}$ are minimized, while ensuring that any combination of antennas can be simultaneously activated. Altogether, this architecture aims to minimize the number of connections to simplify the hardware design.

\item \emph{Architecture $3$. Fully-flexible architecture with minimum losses.} The fully-flexible with minimum connectivity architecture does not guarantee that the insertion losses are minimized, since designing networks with larger $T_{\text{RF}}$ or $T_{\text{AN}}$ might actually reduce the total insertion loss. This counter-intuitive behaviour is illustrated in Table \ref{tab:exampleArchitecture3}, where it can be observed that minimizing the $T_{\text{RF}}$ and $T_{\text{AN}}$ as per the fully-flexible with minimum connectivity architecture does not minimize the total insertion loss. This is because the insertion losses introduced by the basic switches vary depending on their number of ports as per Table \ref{tab:basicSwitches}, an effect particularly noticeable for large $T_{\text{RF}}$ and $T_{\text{AN}}$. In contrast, the fully-flexible architecture with minimum losses selects the $T_{\text{RF}}$ and $T_{\text{AN}}$ that minimize the insertion loss by relaxing \eqref{eq:constraintsFCA2} into
\begin{equation}
T_{\text{RF}} \geq N-M+1, \text{ and } T_{\text{AN}} \geq \min \left( M, T_{\text{RF}} \right).
\label{eq:constraintsFCA3}
\end{equation}
\end{enumerate}

At this point we note that while the different design criteria considered above guarantee full flexibility, their insertion loss can still be as significant as $2$-$3$ dB for massive MIMO systems, as demonstrated in Table \ref{tab:exampleArchitecture3} and shown in Sec.\ \ref{sec:simulationResults}. For this reason, in the following we concentrate on reducing the complexity and losses of RF switching matrices at the expense of reducing their flexibility.

\section{Partially-Connected Switching Architectures and Resulting AS Constraints}
\label{sec:PCSwitching}

The significant insertion loss and complexity of fully-flexible switching matrices in massive MIMO motivate the design of alternative low-complexity switching architectures \cite{7417765}. In this section we consider a partially-connected switching architecture with arbitrary $N$ and $M$ for alleviating the above-mentioned concerns. Specifically, the partially-connected architecture is designed to reduce the number of internal connections and basic switches, i.e.\
\begin{equation}
T_{\text{RF}} = \left\lceil \frac{N}{M} \right\rceil, \text{ and } T_{\text{AN}} = \begin{cases}
1, \mbox{ } \text{ if } \left\lfloor \frac{N}{M} \right\rfloor \geq 2, \\
2, \mbox{ } \text{ otherwise},
\end{cases}
\label{eq:PCPorts}
\end{equation}
where the maximum number of throws per RF chain in the RF-chain switching stage, $T_{\rm RF}$, is defined to guarantee the essential constraint of connecting every antenna to at least one RF chain. Moreover, \eqref{eq:PCPorts} shows that there are two specific cases depending on the ratio $\frac{N}{M}$:

\begin{figure}[!t]
	\centering
		\includegraphics[width=0.38\textwidth]{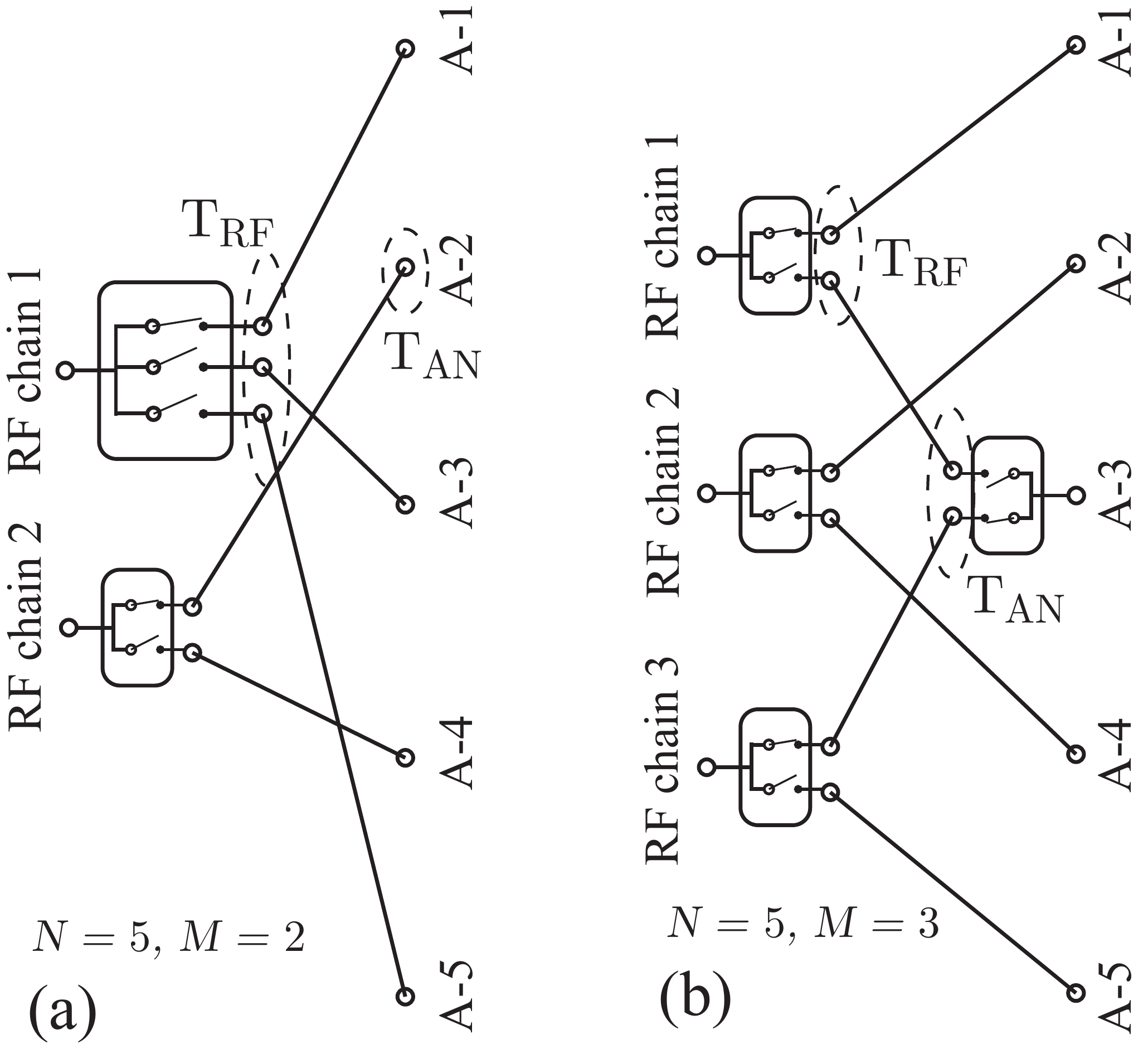}
		\caption{Illustrative block diagrams for the partially-connected switching architectures of (a) $\left\lfloor \frac{N}{M} \right\rfloor \geq 2$ and (b) $\left\lfloor \frac{N}{M} \right\rfloor < 2$.}
\label{fig:examplesConnectivity}
\end{figure}

\begin{enumerate}
\item \emph{Case for $\left\lfloor \frac{N}{M} \right\rfloor \geq 2$.} Partially-connected switching schemes satisfying this condition are those where each antenna is connected to a single RF chain ($T_{\rm AN} = 1$). We note that this architecture comprehends the particular case $N/M =2$ considered in \cite{7417765}, where only SP2T switches are required to ensure antenna connectivity. An illustrative example is shown in \figurename~\ref{fig:examplesConnectivity}(a), where $N = 5$ and $M = 2$. In this architecture, it can be observed that one SP3T must be implemented at the RF-chain switching stage for guaranteeing that every antenna is connected to one RF chain.

\item {\emph{Case for $\left\lfloor \frac{N}{M} \right\rfloor < 2$.}} In these architectures each antenna might be connected to more than one RF chain ($T_{\rm AN} = 2$). Accordingly, employing basic switches SP2T suffices to guarantee connectivity for all $N$ antennas. A specific switching architecture satisfying $\left\lfloor N/M \right\rfloor < 2$ is shown in \figurename~\ref{fig:examplesConnectivity}(b) for $N = 5$ and $M = 3$. This figure shows that the SP2T switches required at the antenna switching stage introduce additional insertion losses with respect to the case $\left\lfloor N/M \right\rfloor \geq 2$.
\end{enumerate}

\subsection{Optimization constraints for AS with Instantaneous CSI}

Overall, the above partially-connected architectures impose a limited flexibility in the AS procedure, since the simultaneous selection of certain antenna combinations is not implementable. In this context, the specific restrictions introduced by the partial connectivity should be considered when defining the antenna selection problem \cite{7417765}. With this purpose, let $\overline{\mathcal{N}} = \left\{ 1, \ldots, N \right\}$ and $\overline{\mathcal{M}} = \left\{ 1, \ldots, M \right\}$ be sets indexing the antennas and the RF chains, respectively. We define the antenna subsets $\mathcal{N}^{i}, i \in \left\{ 1, \ldots, S_{\rm cons} \right\}$ as disjoint sets of antennas sharing at least one common RF chain, i.e.\ $\mathcal{N}^{1} \cup \mathcal{N}^{2} \cup \cdots \cup \mathcal{N}^{S_{\rm cons}} = \overline{\mathcal{N}}$ and $\mathcal{N}^{i} \cap \mathcal{N}^{j} = \varnothing, \forall \mbox{ } i,j \in \left\{ 1, \ldots, S_{\rm cons} \right\}$. Here, $\varnothing$ denotes the empty set and $S_{\rm cons}$ denotes the number of constraints detailed below. Similarly, $\mathcal{M}^{i} \subset \overline{\mathcal{M}}, i \in \left\{ 1, \ldots, S_{\rm cons} \right\}$ represent disjoint sets of RF chains sharing at least one common antenna. For the illustrative example of \figurename~\ref{fig:examplesConnectivity}(b), the sets $\mathcal{M}^{i}$ are defined as $\mathcal{M}^{1} = \left\{ 1, 3 \right\}$, since these RF chains inter-connect antennas $\mathcal{N}^{1} = \left\{ 1, 3, 5 \right\}$, and $\mathcal{M}^{2} = \left\{ 2 \right\}$, which inter-connects antennas $\mathcal{N}^{2} = \left\{ 2, 4 \right\}$. Intuitively, the cardinality of the antenna groups sharing at least one RF chain ($S_{\rm cons}$) also represents the number of constraints required to account for partial connectivity, which is given by
\begin{equation}
S_{\rm cons} = M - N_{\rm ov},
\end{equation}
where $N_{\rm ov}$ represents the number of antennas with the possibility of connecting to more than one RF chain. This is because the partial connectivity architecture imposes an additional constraint for every subset of antennas $\mathcal{N}^{i} \subset \overline{\mathcal{N}}, i \in \left\{ 1, \ldots, S_{\rm cons} \right\},$ interconnected with their corresponding subset of RF chains $\mathcal{M}^{i} \subset \overline{\mathcal{M}}, i \in \left\{ 1, \ldots, S_{\rm cons} \right\}$. For instance, \figurename~\ref{fig:examplesConnectivity}(b) shows that only two out of antennas $\mathcal{N}^{1} = \left\{ 1, 3, 5 \right\}$ can be simultaneously active, since these antennas are only connected to two RF chains ($\vert \mathcal{M}^{1} \vert = 2$). % Similarly, only one antenna out of antennas $\mathcal{N}^{2} = \left\{ 2, 4 \right\}$ can be active because they share a single RF chain ($\vert \mathcal{M}^{2} \vert = 1$).

Without loss of generality, we consider in the following that each RF chain is connected to antennas physically  separated as shown in \figurename~\ref{fig:examplesConnectivity}(a) and \figurename~\ref{fig:examplesConnectivity}(b)\footnote{This consideration is motivated by the results obtained in \cite{7417765} for real propagation environments, where it was shown that inter-connecting a given RF chain to non-adjacent antennas provides a better performance than the connection to adjacent antennas due to their enhanced spatial diversity.}. Based on the above, let us define
\begin{equation}
N_{\rm dist} = \begin{cases}
M, \text{ if } \left\lfloor N/M \right\rfloor \geq 2, \\
N - M, \text{ otherwise}.
\end{cases}
\end{equation}
In plain words, $N_{\rm dist}$ represents the distance between antennas inter-connected to a given RF chain. Similarly, the distance between the RF chains connected to a given subset of antennas $M_{\rm dist}$ is given by
\begin{equation}
M_{\rm dist} = \begin{cases}
1, \text{ if } \left\lfloor N/M \right\rfloor \geq 2, \\
N - M, \text{ otherwise}.
\end{cases}
\end{equation}
For example, \figurename~\ref{fig:examplesConnectivity}(b) shows that RF chains $\mathcal{M}^{1} = \left\{ 1, 3 \right\}$, which inter-connect antennas $\mathcal{N}^{1} = \left\{ 1, 3, 5 \right\}$, are separated by $M_{\rm dist} = N - M = 2$ to minimize the length of the longest wired connection. Accordingly, we can express $\mathcal{N}^{i}$ and $\mathcal{M}^{i}$ for the non-adjacent antenna connectivity considered in this paper as
\begin{equation}
\mathcal{N}^{i}  = \left\{ i, i+N_{\rm dist}, \ldots, i + \left( \left\lceil \frac{N - i + 1}{N_{\rm dist}} \right\rceil - 1 \right) N_{\rm dist} \right\},
\label{eq:setsNi}
\end{equation}
where each RF chain connects to antennas with indices separated by $N_{\rm dist}$ and 
\begin{equation}
\mathcal{M}^{i}  = \left\{ i, i + M_{\rm dist}, \ldots, i + \left( \left\lceil \frac{M - i + 1}{M_{\rm dist}} \right\rceil  - 1 \right) M_{\rm dist} \right\},
\label{eq:setsMi}
\end{equation}
where each antenna connects to RF chains with indices separated by $M_{\rm dist}$. Considering the above, the AS optimization problem for general partially-connected architectures can be formulated as
\begin{subequations}
\begin{align}
& \mathcal{P}_{2}: \underset{\mathbf{S}, \mathbf{P}}{\text{ maximize }} \log_{2} \det \left( \mathbf{I}_{K} + \rho \mathbf{P} \mathbf{H} \mathbf{S} \mathbf{H}^{\rm H}  \right) \label{eq:optimizationPC} \\
& \text{ subject to } \sum_{i = 1}^{N} \mathbf{S}_{i,i} = M,  \\
& \kern 4.6em 0 \leq \mathbf{S}_{i,i} \leq  1, \mbox{ } \forall i \in \left\{ 1, \ldots , N \right\},  \\
& \kern 4.3em \sum_{j \in \mathcal{N}^{i}} \mathbf{S}_{j,j} =  \vert \mathcal{M}^{i} \vert ,\mbox{ } \forall \mbox{ } i \in \left\{ 1, \ldots, S_{\rm cons} \right\}, \label{eq:PCConstraintP5}
\end{align}
\end{subequations}
In resemblance with \cite{7172496,7417765, 1687757}, $\mathcal{P}_{2}$ is solved by first optimizing over $\mathbf{S}$, while setting $\mathbf{P} = \mathbf{I}_{K}$. Note that the resultant convex optimization problem is produced by relaxing binary constraints, so the solution $\mathbf{\widetilde{S}}^{\star}$ can be obtained by choosing the $M$ largest entries of $\mathbf{S}$ compliant with the partial connectivity limitations. Subsequently, the user power allocation matrix $\mathbf{\widetilde{P}}^{\star}$ is computed by maximizing \eqref{eq:capacityDL} with $\mathcal{M}$ fixed as per \eqref{eq:selectionMatrixDefinition}. This is also a convex optimization problem with a waterfilling-type solution \cite{1237143}.

\subsection{Optimization constraints for AS with Power-Based CSI}

The specific RF switching architecture selected also influences the procedure for performing AS based on the channel power at each of the antenna ports. Indeed, the optimization problem $\mathcal{P}_{1}$ must incorporate additional constraints due to the partial connectivity. As a result, the optimization problem for power-based AS under limited connectivity can be expressed as
\begin{subequations}
\begin{align}
\mathcal{P}_{3}: & \underset{\mathbf{S}}{\text{ maximize }} \overline{\mathbf{h}} \mathbf{S}, \label{eq:PCPBASOpt} \\
& \text{ subject to } \sum_{i = 1}^{N} \mathbf{S}_{i,i} = M, \\
& \kern 4.65em   \mathbf{S}_{i,i} \in \left\{ 0, 1 \right\}, \mbox{ } \forall \mbox{ } i \in \left\{ 1, \ldots , N \right\}, \\
& \kern 4.3em \sum_{j \in \mathcal{N}^{i}} \mathbf{S}_{j,j} =  \vert \mathcal{M}^{i} \vert, \mbox{ } \forall \mbox{ }  i \in \left\{ 1, \ldots, S_{\rm cons} \right\},  \label{eq:PCConstraintP6}
\end{align}
\end{subequations}
where the sets $\mathcal{N}^{i}$ and $\mathcal{M}^{i}$ are defined in \eqref{eq:setsNi} and \eqref{eq:setsMi}, respectively. Intuitively, the solution of $\mathcal{P}_{3}$ can be obtained by selecting the $\vert \mathcal{M}^{i} \vert$ largest entries of $\overline{\mathbf{h}}_{\mathcal{N}^{i}}, \forall \mbox{ }  i \in \left\{ 1, \ldots, S_{\rm cons} \right\}$.

\subsection{Implications of Reducing Connectivity}

The constraints introduced to account for partial connectivity have a number of practical system-level implications that we detail as follows:

\begin{itemize}
\item \emph{Insertion losses.} The insertion losses introduced will be smaller in the design with partial switching connectivity due to the smaller number of basic switches required to implement the switching matrix. For instance, in the example of Table \ref{tab:exampleArchitecture3}, the insertion losses are reduced from $L \approx 3$ dB to $L = 0.5$ dB.

\item \emph{Power-based AS.} The need of solving $\mathcal{P}_{3}$ entails that the overall signal processing complexity of the power-based AS under restricted RF switching connectivity is modified w.r.t.\ its fully-flexible counterpart. This is because of the reduced length of the sub-arrays from from which the largest elements are selected (from $N$ to $\max \left( \vert \mathcal{M}^{i} \vert \right) \forall \mbox{ }  i \in \left\{ 1, \ldots, S_{\rm cons} \right\}$).

\item \emph{System performance.} Limiting the connectivity affects the number of possible antenna combinations that can be selected. Due to its importance, this aspect is studied in Sec.\ \ref{sec:performanceAnalysis}, where the ergodic sum capacity loss w.r.t.\ fully-flexible architectures is characterized for the case of power-based AS.

\item \emph{Baseband signal processing.} The digital signal processor (DSP) must account for the reduced connectivity in order to perform the precoding / detection operations. This is because the order of the antenna channels might be different from the conventional order at the antenna ports. Let us consider the architecture shown in \figurename~\ref{fig:examplesConnectivity}(a) and assume that antennas $\mathcal{N}^{2} = 2$ and $\mathcal{N}^{1} = 5$ are those that maximize the sum capacity. For adequate symbol-to-antenna mapping, the DSP should be aware that the first RF chain will be connected to antenna port $5$ whereas the second RF chain will be connected to antenna port $2$. For instance, the precoding/detection weight applied to the first RF chain should be that of antenna $5$ to guarantee a correct system operation.

\item \emph{Analog hardware complexity.} The employment of a reduced number of basic switches and connections can reduce cross-coupling between hardware components that are physically close, the time required for calibrating the input-output transfer function and the overall economic cost for implementing the switching matrix. 
\end{itemize}

\subsection{Practical Hardware Implementation for Power-Based AS in massive MIMO}

\begin{figure}[!t]
	\centering
		\includegraphics[width=0.45\textwidth]{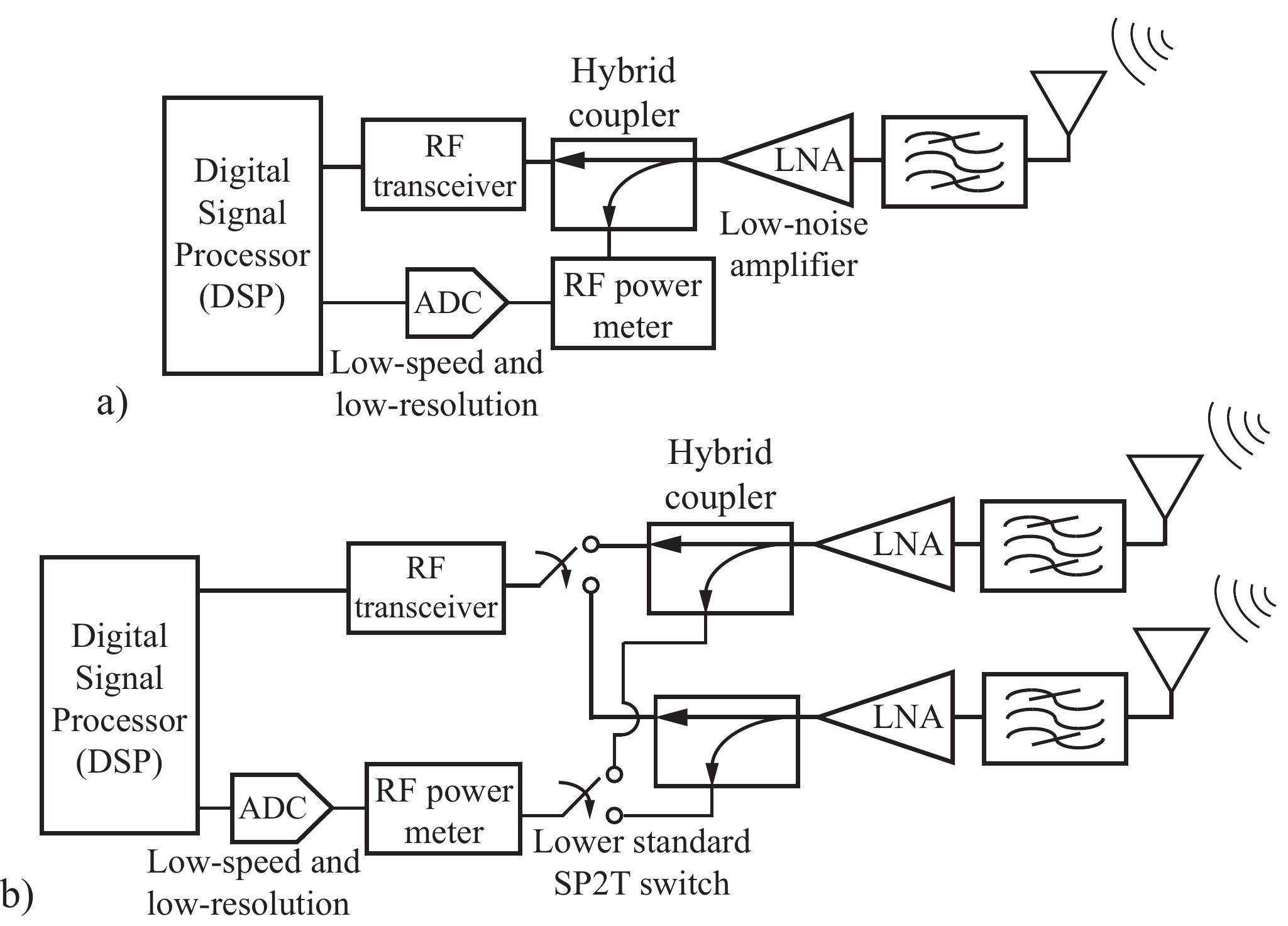}
		\caption{Block diagram of different hardware implementations for measuring channel power. (a) One parallel power meter chain per antenna and (b) limited number of power meter chains with additional selection switches.}
\label{fig:PBArchitectures}
\end{figure}

The solution advocated in \cite{4633647} for reducing the CSI acquisition time in power-based AS is based on implementing analog power estimators at each antenna as shown in \figurename~\ref{fig:PBArchitectures}(a), i.e.\ $N_{\rm PM} = N$, where $N_{\rm PM}$ refers to the number of power meters. Note that here we consider that the speed and resolution requirements of the analog-to-digital converters (ADC) required in the parallel power-meter chains can be relaxed because power-based AS solely relies on the relative differences between the received signals from different antennas. Nevertheless, additional data ports are required at the digital signal processor (DSP) to acquire this data.

The excessive number of antennas implemented in massive MIMO systems also motivates the implementation of a considerable number of RF chains, even when AS is implemented \cite{7172496}. Since each RF chain captures the channel power information during the uplink data stage, this entails that the solution illustrated in \figurename~\ref{fig:PBArchitectures}(a) acquires a significant amount of redundant data. This redundancy is eliminated in the architecture considered in \figurename~\ref{fig:PBArchitectures}(b), where only $N_{\rm PM} = N-M$ power meters are required. The particular example shown in \figurename~\ref{fig:PBArchitectures}(b) corresponds to an architecture with $N = 2M$. When compared with the scheme of \figurename~\ref{fig:PBArchitectures}(a), it can be seen that, while additional RF switches are required, the number of parallel power-meter chains and data ports at the DSP can be substantially reduced.

\section{Performance Analysis: Degradation due to Limited Switching Connectivity}
\label{sec:performanceAnalysis}

The partial connectivity architecture presented in Sec.\ \ref{sec:PCSwitching} reduces the insertion loss introduced by the switching matrix at the expense of limiting the number of number of antenna combinations that can be simultaneously selected, which entails a loss in performance. In this section, we characterize this loss by following the intuitive approach adopted in \cite{4600226} for power-based AS. The reasons for focusing on the simpler power-based AS are twofold: a) its reduced CSI acquisition time as detailed in Sec.\ \ref{sec:CSIAcquisition} by use of RF power meters, and b) the negligible performance loss introduced by this scheme when compared with the selection based on instantaneous CSI for the channels considered in this paper.

\subsection{Ergodic Capacity Approximation for Fully-Flexible AS}

The analysis developed in \cite{4600226} essentially relies on approximating $\mathbf{H}_{[ \mathcal{M} ]}$ in \eqref{eq:capacityDL} by a matrix $\mathbf{V} \in \mathbb{C}^{K \times M}$ with entries following an identical distribution but with each column having different variances, i.e.\
\begin{equation}
\mathbf{H}_{[ \mathcal{M} ]} \approx \mathbf{V} = \mathbf{G} \boldsymbol\Theta,
\label{eq:channelApprox}
\end{equation}
where $\mathbf{G} \in \mathbb{C}^{K \times M}$ is a matrix whose entries follow the same distribution of those from $\mathbf{H}$ and $\boldsymbol\Theta \in \mathbb{R}^{M \times M}$ is a diagonal matrix whose definition is considered in the following. Specifically, let ${B}_{c} = \sum_{i}^{K} \vert h_{c,i} \vert^{2} = \norm{\mathbf{h}_{c}}^{2}$ denote the norm of the $c$-th column of $\mathbf{H}$ and define $B_{t:N}$ as the $t$-th smallest column norm of $\mathbf{H}$ as per $\mathcal{W} = \left\{ B_{1:N} < B_{2:N} < \cdots < B_{N:N} \right\}$. Then, the diagonal entries of $\boldsymbol\Theta$ are given by \cite{4600226}
\begin{equation}
\theta_{i,i} = \mathbb{E} \left[ B_{t_{i}:N} \right] / \sqrt{K}, \mbox{ } t_{i} \in \left\{ 1, \ldots, N \right\}.
\label{eq:diagonalThetaDefinition}
\end{equation}
The definition of the indices $t_{i}$ is straightforward and deterministic for the case of fully-flexible switching networks, since the power-based AS will always select the antennas corresponding to the largest column norms of $\mathbf{H}$, i.e.
\begin{equation}
\boldsymbol\Theta = \diag \left( \sqrt{B_{N:N}}, \sqrt{B_{N-1:N}}, \ldots, \sqrt{B_{N-M+1:N}} \right) / \sqrt{K}.
\label{eq:ThetaDefinitionFC}
\end{equation}

A simpler approximation of the resultant channel in \eqref{eq:channelApprox} can be obtained by averaging the power scaling factors of the selected ordered statistics in \eqref{eq:diagonalThetaDefinition}, which for fully-flexible switching networks yields
\begin{equation}
\widetilde{P}_{\rm FF} = \frac{1}{KM} \sum_{i=1}^{M} \mathbb{E}_{\mathbf{H}} \left[ B_{N-i+1:N} \right],
\label{eq:powerScalingFactor}
\end{equation}
where $\mathbb{E}_{\mathbf{H}}$ denotes that the expectation is taken over the small-scale fading parameters of the random channel $\mathbf{H}$. In contrast with \cite{4600226}, in this work we rely on an approximation of the first moment of the $t$-th ordered random variable $B_{t:N}$. This is because the expressions employed in \cite{4600226} are not adequate for systems with a large number of antennas due to the excessive complexity involved in their calculation. Specifically, we approximate the first moment of the $t$-th ordered random variable $B_{t:N}$ in uncorrelated Rayleigh flat-fading channels as \cite{gupta1960order, breiter1967tables}
\begin{equation}
\mathbb{E}_{\mathbf{H}} \left[ B_{t:N} \right]  \approx \frac{c}{2} \frac{N! \left( \left(K-1\right)! \right)^{-1}}{ \left(t-1\right)! \left(N-t\right)!} \int_{-1}^{1} f(x) \left( x + 1 \right) \mbox{ }dx,
\label{eq:expectationOrderedRVRayleigh}
\end{equation}
where $f(x)$ is given by
\begin{equation}
f(x) = \left( 1 - \sum_{j = 0}^{K - 1} \frac{e^{-y} y^{j}}{j!} \right)^{k-1} \left(  \sum_{j = 0}^{K - 1} \frac{e^{-y} y^{j}}{j!} \right)^{N - k} e^{-y} y^{K}.
\end{equation}
In the above expressions, $y \triangleq c (x + 1)^2 / 4$ and $c$ is a large constant ($c \approx 100$ provides sufficient accuracy in the approximation \cite{breiter1967tables}). The result of the definite integral in \eqref{eq:expectationOrderedRVRayleigh} can be easily computed via Gauss-Jacobi quadrature  \cite{gautschi2011numerical}.

The resultant ergodic capacity with the power scaling approximation can be subsequently expressed as \cite{4600226}
\begin{equation}
C_{\rm PS-FF} = \mathbb{E}_{\mathbf{G}} \left[ \log_{2} \det \left( \mathbf{I}_{K} + \rho \widetilde{P}_{\rm FF} \mathbf{G} \mathbf{G}^{\rm H}  \right) \right],
\label{eq:CPSApprox}
\end{equation}
where the expectation over $\mathbf{G}$ has been analytically derived for a variety of correlated and uncorrelated communication channels \cite{4600226}.

\subsection{Ergodic Capacity Approximation for Partially-Connected AS}

While the approximation in \eqref{eq:CPSApprox} accurately characterizes the ergodic capacity attainable by fully-flexible schemes, an optimal power-based AS cannot be guaranteed under limited connectivity in general. This is because selecting the antenna combination with the largest column norms of $\mathbf{H}$ might not be feasible due to the additional constraints imposed in the optimization problem. This entails that \eqref{eq:ThetaDefinitionFC} and \eqref{eq:powerScalingFactor} are no longer valid for partially-connected networks. For this reason we propose to further approximate \eqref{eq:powerScalingFactor} by taking into consideration the probability of selecting a given subset of the ordered set $\mathcal{W} = \left\{ B_{1:N} < B_{2:N} < \cdots B_{N:N} \right\}$. 

With the above purpose, let us define  $\mathcal{B}^{j}, j \in \left\{ 1, \ldots, \binom {N}{M} \right\}$ of cardinality $\vert \mathcal{B}^{j} \vert = M$ as discrete sets containing a given combination of ordered column norms of $\mathbf{H}$. For instance, in the architecture shown in \figurename~\ref{fig:examplesConnectivity}(a), the relevant sets $\mathcal{B}^{j}$ with non-zero probability of being selected following a power-based criterion are $\mathcal{B}^{1} = \left\{ 1, 2 \right\}$, $\mathcal{B}^{2} = \left\{ 1, 3 \right\}$ and $\mathcal{B}^{3} = \left\{ 1, 4 \right\}$. Note that the antenna with the largest power can always be selected even under partial connectivity restrictions. Moreover, let $T_{j} \in \left\{ 0, 1 \right\}$  be a binary random variable that determines whether the specific combination of columns of $\mathbf{H}$ determined by $\mathcal{B}^{j}$ is selected or not. Intuitively, the limited connectivity restricts a given combination of antennas $\mathcal{B}^{j}$ to be selected with a given probability. This entails that, in contrast with \eqref{eq:powerScalingFactor}, the expectation in \eqref{eq:diagonalThetaDefinition} must also be taken with respect to the discrete random variables $T_{j}$. Therefore, we propose to theoretically approximate the performance of power-based AS for partially-connected switching architectures by employing
\begin{equation}
\widetilde{P}_{\rm PC} = \frac{1}{KM} \mathbb{E}_{T_{j}} \left[ \sum_{i=1}^{M} \mathbb{E}_{\mathbf{H}} \left[ B_{\mathcal{B}^{j}_{i}  :N} \right] \right], 
\label{eq:PSPartialConnectivity}
\end{equation}
where $\mathcal{A}_{i}$ refers to the $i$-th entry of a set $\mathcal{A}$ and the outer expectation is taken over the set of random discrete variables $T_{j}$. Closed-form expressions for the inner expectation in \eqref{eq:PSPartialConnectivity} are already available for multiple channels such as those with Rayleigh flat-fading as per \eqref{eq:expectationOrderedRVRayleigh} \cite{4600226}. Instead,  the outer expectation corresponds to that of a discrete random variable, which is given by
\begin{align}
\widetilde{P}_{\rm PC} & = \frac{1}{KM} \sum_{j=1}^{\binom {N}{M}}\left( \sum_{i=1}^{M}  \mathbb{E}_{\mathbf{H}} \left[ B_{\mathcal{B}^{j}_{i}  :N} \right] \right) \times P \left( T_{j} \right) \nonumber \\
& =  \sum_{j=1}^{\binom {N}{M}} \left( \widehat{P}_{j} \times P \left( T_{j} \right) \right),
\label{eq:powerScalingPC}
\end{align}
where $P \left( T_{j} \right)$ refers to the probability of jointly selecting the ordered channel columns determined by $\mathcal{B}^{j}$. The probabilities associated with selecting $\mathcal{B}^{j}$ can be computed depending on the specific switching connectivity and the particular channel characteristics. Specifically, by following the chain rule of probability we have
\begin{align}
P  \left( T_{j} \right) & = P  \left( \bigcap_{i=1}^{M} \text{ selecting } \mathcal{B}^{j}_{i} \right)   \nonumber \\ & = \prod_{i =1}^{M} P  \left( \text{ selecting } \mathcal{B}^{j}_{i} \vert \bigcap_{r=1}^{i-1} \text{ selected } \mathcal{B}^{j}_{r} \right),
\label{eq:jointProbabilityFormula}
\end{align}
where $\cap$ denotes the intersection of events. To simplify the derivation of the joint probabilities in \eqref{eq:jointProbabilityFormula}, in the following we concentrate on channels $\mathbf{H}$ adopting the form \cite{6375940}
\begin{equation}
\mathbf{H} = \mathbf{R}^{\frac{1}{2}} \mathbf{F}, 
\end{equation}
where $\mathbf{R} \in \mathbb{C}^{K \times K}$ is the deterministic channel covariance matrix and $\mathbf{F} \in \mathbb{C}^{K \times N}$  is a matrix of independent and identically distributed (i.i.d.) random variables following $\mathbf{f}_{i,j} \sim\mathcal{C}\mathcal{N}(0,1)$.

The above assumption entails that the probability of finding the $t$-th ordered statistic $B_{t:N}, t \in \left\{ 1, \ldots, N\right\}$ at a given antenna is equal for all antenna elements $\mathcal{N}$, which makes computing the probabilities in \eqref{eq:jointProbabilityFormula} straightforward for a given switching connectivity. An example of this derivation for $N = 5$ and $M = 2$ is provided in Appendix A for completeness. Once the probabilities of selecting a given ordered channel column combination $P  \left( T_{j} \right)$ in \eqref{eq:powerScalingPC} are determined, an approximation to the analytical ergodic capacity of AS systems with partially-connected switching matrices can be expressed as
\begin{equation}
C_{\rm PS-PC} \approx \mathbb{E}_{\mathbf{G}} \left[  \log_{2} \det \left( \mathbf{I}_{K} + \rho \widetilde{P}_{\rm PC} \mathbf{G} \mathbf{G}^{\rm H}  \right) \right].
\label{eq:finalApproximation1}
\end{equation}

An alternative approximation to the ergodic capacity of partially-connected switching matrices can be derived by directly computing the expectation of the capacity over both the discrete random variables $T_{j}$ and $\mathbf{G}$. In this particular case, the approximation of the ergodic capacity is given by
\begin{equation}
C_{\rm PS-PC} \approx \sum_{j=1}^{\binom {N}{M}}  \mathbb{E}_{\mathbf{G}} \left[  \log_{2} \det \left( \mathbf{I}_{K} + \rho \widehat{P}_{j} \mathbf{G} \mathbf{G}^{\rm H} \right) \right] \times P \left( T_{j} \right),
\label{eq:finalApproximation2}
\end{equation}
where $\widehat{P}_{j}$ is defined in \eqref{eq:powerScalingPC}. 

\begin{figure}[!t]
	\centering
		\includegraphics[width=0.45\textwidth]{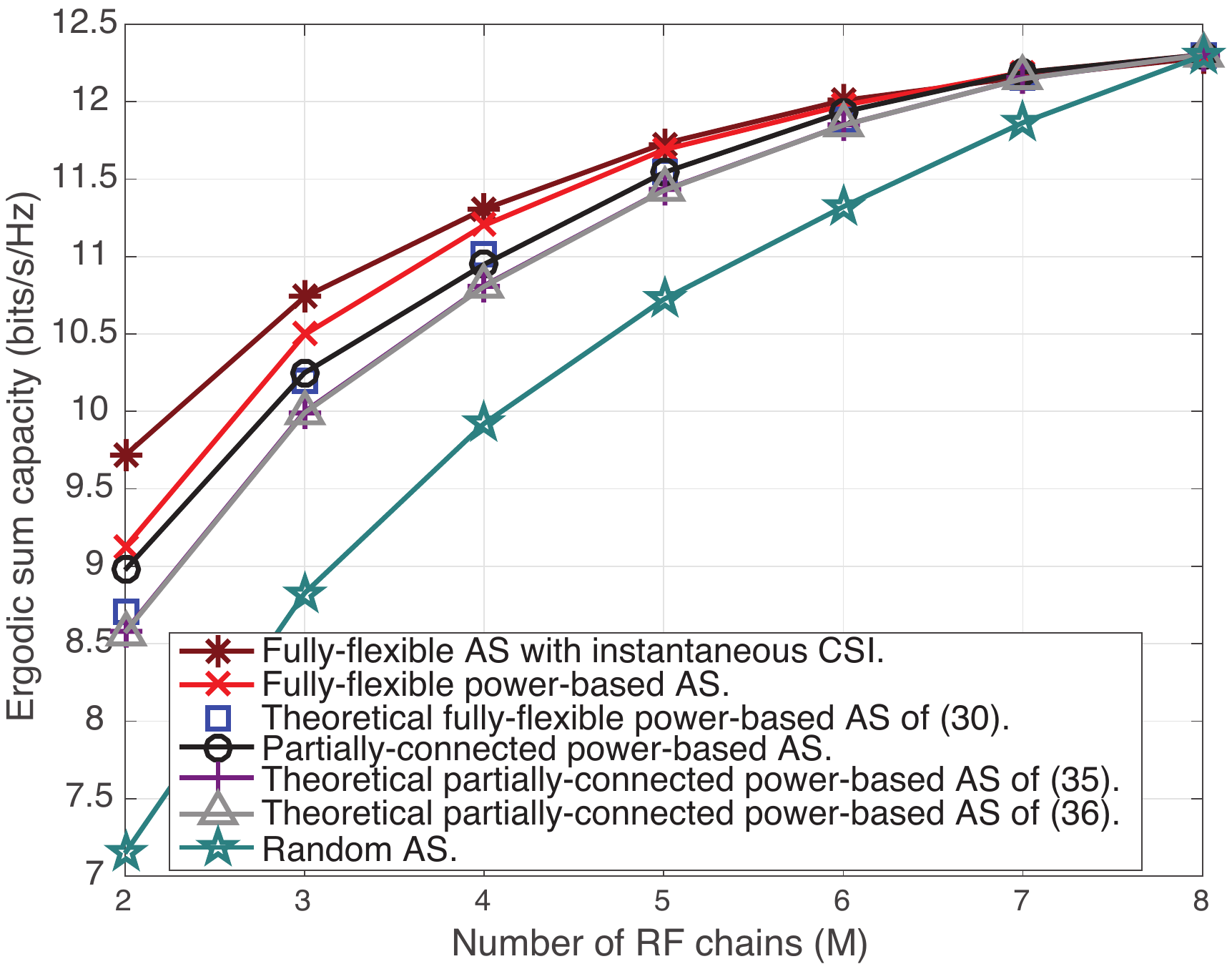}
		\caption{Theoretical and simulated ergodic capacities (bits/s/Hz) vs.\ $M$. $N = 8$ for AS systems and $N = M$ for MU-MIMO without AS, $K = 2$ and $\rho = 10$ dB. CSI acquisition overheads and insertion losses are not considered.}
\label{fig:theoreticalResult}
\end{figure}

The accuracy of the derived approximations can be observed in \figurename~\ref{fig:theoreticalResult}, which represents the ergodic capacity attained by the schemes considered in this paper against increasing $M$ for a system with $N = 8$ for AS systems and $N = M$ for conventional MIMO without AS, $K = 2$ and $\rho = 10$ dB. Here we consider a small-scale MIMO setup for illustrative reasons since, as shown in Sec.\ \ref{sec:simulationResults}, the differences between the switching architectures vanish for large $N$. For the results of this figure, we consider an uncorrelated Rayleigh flat-fading channel, ignore the overheads associated with CSI acquisition and assume an identical transmission power at the antenna ports, i.e., the switching insertion losses are precompensated. In spite of this unfavourable assumption for the power-based criterion, it can be seen that it approaches the performance attained by instantaneous CSI-based AS for different values of $M$, an observation consistent with previous works \cite{4600226,1341263}. \figurename~\ref{fig:theoreticalResult} also shows that the switching architecture with partial connectivity only experiences a slight performance loss w.r.t.\ the fully-flexible scheme, which is also coherent with the results obtained for $N = 2M$ in the real propagation channels of \cite{7417765}. The slight performance loss can be explained by noting that, in general, there exists a large probability of selecting antennas with significant channel powers, even if these are not strictly the largest ones as shown in Appendix A. This result combined with those obtained in Sec.\ \ref{sec:simulationResults} motivate the employment of power-based AS with partial connectivity, especially for massive MIMO where the insertion losses introduced by the switching matrices can be significant. Moreover, \figurename~\ref{fig:theoreticalResult} shows that the proposed approximations in \eqref{eq:finalApproximation1} and \eqref{eq:finalApproximation2} capture the performance loss produced by the partial connectivity architecture.

\section{Energy Efficiency}
\label{sec:EEAnalysis}

The enhancement of the communication's energy efficiency is a key driver for considering AS implementations in massive MIMO due to the possibility of reducing the number of RF chains simultaneously active \cite{7172496,7417765,6666272}. This comes, however, at the expense of introducing additional insertion losses in the RF switching stage, hence posing an energy efficiency trade-off that we aim to analyze in this paper. The energy efficiency for downlink transmission can be expressed as \cite{6666272,7031971,6731024}
\begin{equation}
\xi = \frac{R_{\rm sum}}{P_{\rm PA} + P_{\rm RF} + M \left( P_{\rm ADC} + P_{\rm DAC} + P_{\rm int} \right) + P_{\rm BB}},
\label{eq:EEFormula}
\end{equation}
where $R_{\rm sum}$ (bits/s) refers to the effective downlink sum rates after accounting for the CSI acquisition and uplink overheads, and $P_{\rm PA} = \frac{P_{\rm t}}{\kappa}$ is the power consumed by power amplifiers (PAs) with efficiency $\kappa$ to produce an output signal of $P_{\rm t}$ Watts. Moreover, $P_{\rm RF} = M  P_{\rm CIR} + P_{\rm LO}$ represents the power consumed by the analog hardware components excluding the PAs. Here, $P_{\rm CIR}$ is the power consumed by the analog circuitry required in each of the RF chains, and $P_{\rm LO}$ denotes the power consumed by the local oscillator. $P_{\rm ADC}$ and $P_{\rm DAC}$ refer to the power consumption of dual channel (I-Q) analog-to-digital converters (ADCs) and digital-to-analog converters (DACs), respectively, whereas $P_{\rm BB}$ and $P_{\rm int}$ represent the power of the DSP baseband signal processing and its data interface per RF chain. The baseband power consumption can be decomposed as $P_{\rm BB} = p_{\rm c} C$, where $p_{\rm c}$ characterizes the power consumed per floating point operation and $C = C^{\rm tr}_{\rm corr} + C^{\rm dl}_{\rm data}$ includes the number of floating point operations for pilot correlation in the CSI acquisition stage $\left( C^{\rm tr}_{\rm corr} \right)$ and for precoding in the downlink data transmission stage $\left( C^{\rm dl}_{\rm data} \right)$. Additionally, the power consumed per RF chain by the DSP data interface is given by
\begin{equation}
P_{\rm int} = p_{\rm int} \left( 2 \cdot S_{\rm ADC} \cdot b_{\rm ADC}  + 2 \cdot S_{\rm DAC} \cdot b_{\rm DAC} \right),
\label{eq:interfacePowerConsumption}
\end{equation}
where $p_{\rm int}$ denotes the power consumption per Gbps, $b_{\rm ADC}$ and $b_{\rm DAC}$ refer to the resolution in bits of the ADCs and DACs, while $S_{\rm ADC}$ and $S_{\rm DAC}$ represent their sampling rate per port (I-Q) in Gbps. The values considered for the above parameters can be found in Table \ref{tab:powerConsumption}, where $N_{\rm coh}$ denotes the number of channel coherence blocks per second and a system operating below 6 GHz has been considered.

\small
\begin{table}[!t] 
\begin{center}
{\renewcommand{\arraystretch}{1.25}
\caption{Energy efficiency parameters}
\begin{tabular}{| p{3.53cm} | p{4.43cm}|} 
 \hline
{\bf \emph{Parameter}} & {\bf \emph{Characteristics}}
\\ \hline \hline
$P_{\rm ADC} = 233$ mW \cite{ADC} & $b_{\rm ADC}$ = 12 bits, $S_{\rm ADC} = 125$ MSPS
\\ \hline
$P_{\rm DAC} = 232$ mW \cite{DAC} & $b_{\rm DAC}$ = 14 bits, $S_{\rm DAC} = 125$ MSPS
\\ \hline
$p_{\rm int} = 25$ mW/Gbps \cite{erdmann2015heterogeneous} & Parallel interface LVDS
\\ \hline
$P_{\rm cir} = 1$ W \cite{7031971} & -
\\ \hline
$P_{\rm LO} = 2$ W \cite{7031971} &  - 
\\ \hline
 $p_{\rm c}^{-1} = 12.8$ Mflops / mW \cite{7031971} & Eight-core C6678 DSP
\\ \hline
 $\kappa = 0.39$ \cite{7031971} & -
 \\ \hline
 $P_{\rm t} = 46$ dBm \cite{ghosh2010fundamentals} & LTE macro BS
\\ \hline \hline 
{\bf \emph{Operation}} & {\bf \emph{Computational complexity}}
\\ \hline
$C^{\rm tr}_{\rm corr}$ \cite{6415389} & $8 N_{\rm coh} \eta_{\rm tr} M K$ flops
\\ \hline
$C^{\rm dl}_{\rm data}$ \cite{7109850,6415389} & $N_{\rm coh} \left( 4 K^{2} M + \eta_{\rm dl} \left( 8 K M \right) \right)$ flops
\\ \hline
\end{tabular}
\label{tab:powerConsumption}}
\end{center}
\end{table}\normalsize

To explicitly account for the impact of the insertion losses on the system's energy efficiency, in this work we incorporate $L$ directly in $P_{\rm t}$ above, that is, we consider that the RF switching matrix is placed after the PAs in the transmission chain \cite{1284943,1227919}. This entails that PAs in systems with AS will be required to produce output signals with larger power to compensate for the insertion loss of the switching stage, i.e.\ $P^{\rm AS}_{\rm t} = P^{\rm no-AS}_{\rm t} \times L$ \cite{1284943,1227919}. While other hardware solutions such as placing the switching matrices before the PAs are certainly feasible, considering the impact of the insertion losses on the resultant system's energy efficiency also becomes more intricate. This is because precise knowledge of the PA response, which is non-linear and component-dependent, would be required to quantify the additional power required per RF chain.

\section{Simulation Results}
\label{sec:simulationResults}

In this section we present numerical results for characterizing the performance of the switching architectures considered in this paper. In particular, we consider AS schemes based on both power-based and instantaneous CSI decisions under uncorrelated Rayleigh flat-fading channels. Similarly to \cite{1284943, 1341263, 6725592, 7247768}, we average over channel realizations and, unless otherwise stated, consider a single carrier transmission. Moreover, we show the sum rates attained via dirty paper coding (DPC), which attains the system's ergodic sum capacity, and for the more practical ZF precoder, computed as per \cite{7172496, 7417765}. The results are obtained for massive MIMO systems, since the insertion losses and complexity of the fully-flexible switching networks are critical in these systems due to both the large number of RF chains and antennas deployed. For simplicity, in the following we consider that the power losses introduced by the switching matrices correspond to those of the input-output combination with largest insertion loss, i.e.\ the critical signal path.

\begin{figure}[!t]
	\centering
		\includegraphics[width=0.45\textwidth]{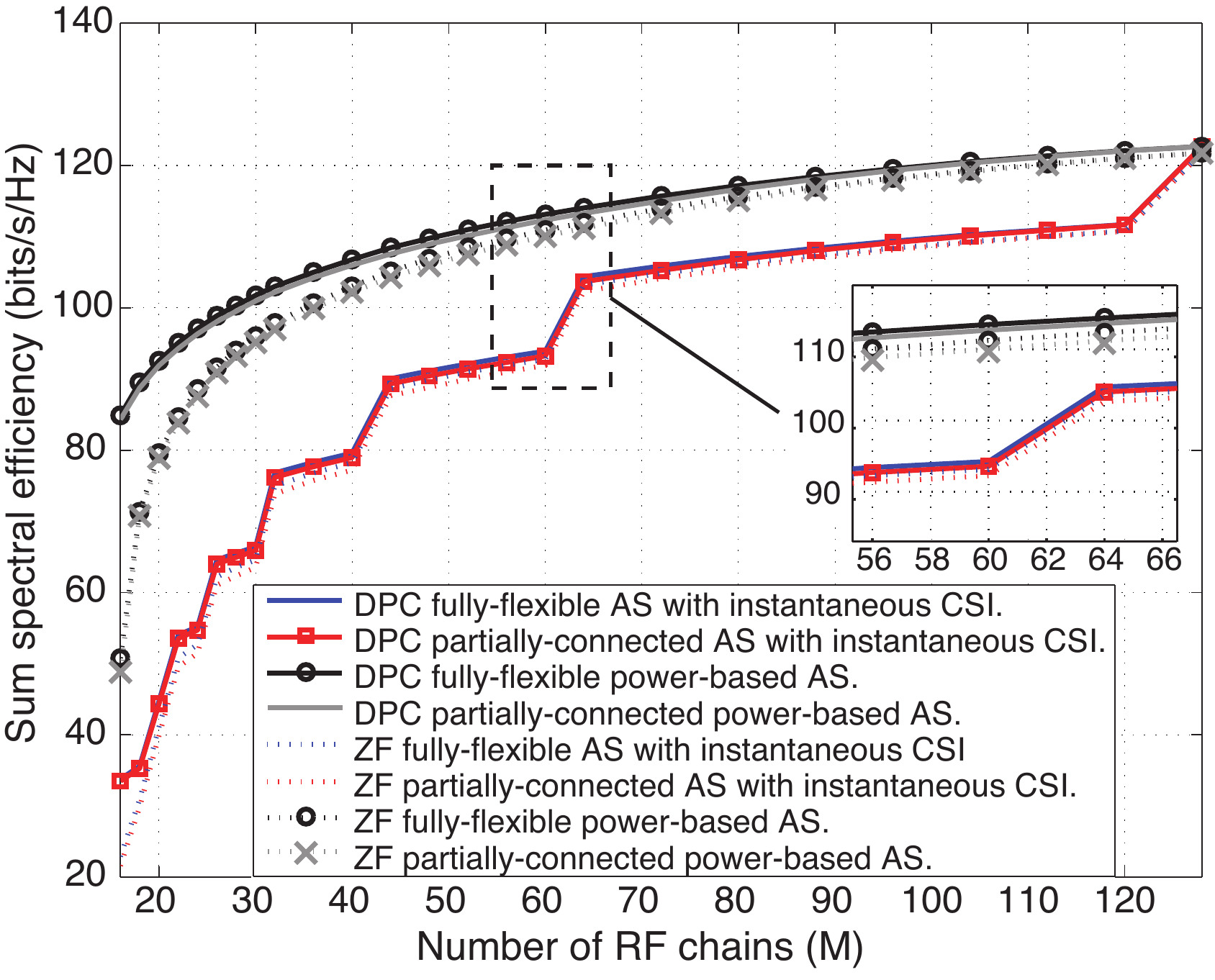}
		\caption{Sum spectral efficiency (bits/s/Hz) vs.\ $M$ for DPC and ZF. $N = 128$, $K = 16$, $\rho = 15$ dB, $\eta_{\rm coh} = 200$ and $\eta_{\rm dl} = 0.7 \times \lceil \left( \eta_{\rm coh} - \eta_{\rm tr} \right) \rceil$. CSI acquisition overheads are considered, whereas insertion losses are ignored.}
\label{fig:spectralEfficiencyCSICost}
\end{figure}

The results of \figurename~\ref{fig:spectralEfficiencyCSICost} represent the sum spectral efficiency of the schemes considered in this paper against increasing number of RF chains $M$ in a system with $N = 128$, $K = 16$ and $\rho = 15$ dB. This figure concentrates on the impact of having a limited channel coherence interval and the need for training in AS by a) ignoring the losses associated with the switching matrices and b) considering $\eta_{\rm coh} = 200$. This channel coherence interval approximates a fast-varying communication channel as per standard LTE OFDM parameters, where it is considered that the channel remains approximately constant throughout $14$ orthogonal subcarriers of $15$ kHz and $14$ OFDM symbols occupying $1$ ms \cite{6415389,7402270}. This value could correspond to a system designed to operate at high frequencies and/or to serve users with speeds of up to 500 km/h at $2$ GHz, such as LTE \cite{6415389,7402270}. Without loss of generality, we consider that $70\%$ of the remaining time-frequency resources after channel estimation are allocated to downlink transmission, i.e.\ $\eta_{\rm dl} = 0.7 \times \lceil \left( \eta_{\rm coh} - \eta_{\rm tr} \right) \rceil$. \figurename~\ref{fig:spectralEfficiencyCSICost} critically shows that full CSI acquisition becomes suboptimal and power-based AS becomes a more attractive approach when a short channel coherence interval is considered.  This is because acquiring instantaneous CSI to perform the selection requires a larger training period, which results in a performance loss particularly significant for reduced $M$. Instead, power-based AS can employ the $N_{\rm PM} = N - M$ power meters integrated in the analog stage for estimating the channel powers and, exploiting channel reciprocity, for performing the selection without requiring accurate CSI for all antennas. The results of \figurename~\ref{fig:spectralEfficiencyCSICost} also demonstrate sudden rate variations for increasing $M$ in the instantaneous CSI-based AS due to the need of multiplexed training as per \eqref{eq:channelAcquisitionRequired}.

\begin{figure}[!t]
	\centering
		\includegraphics[width=0.45\textwidth]{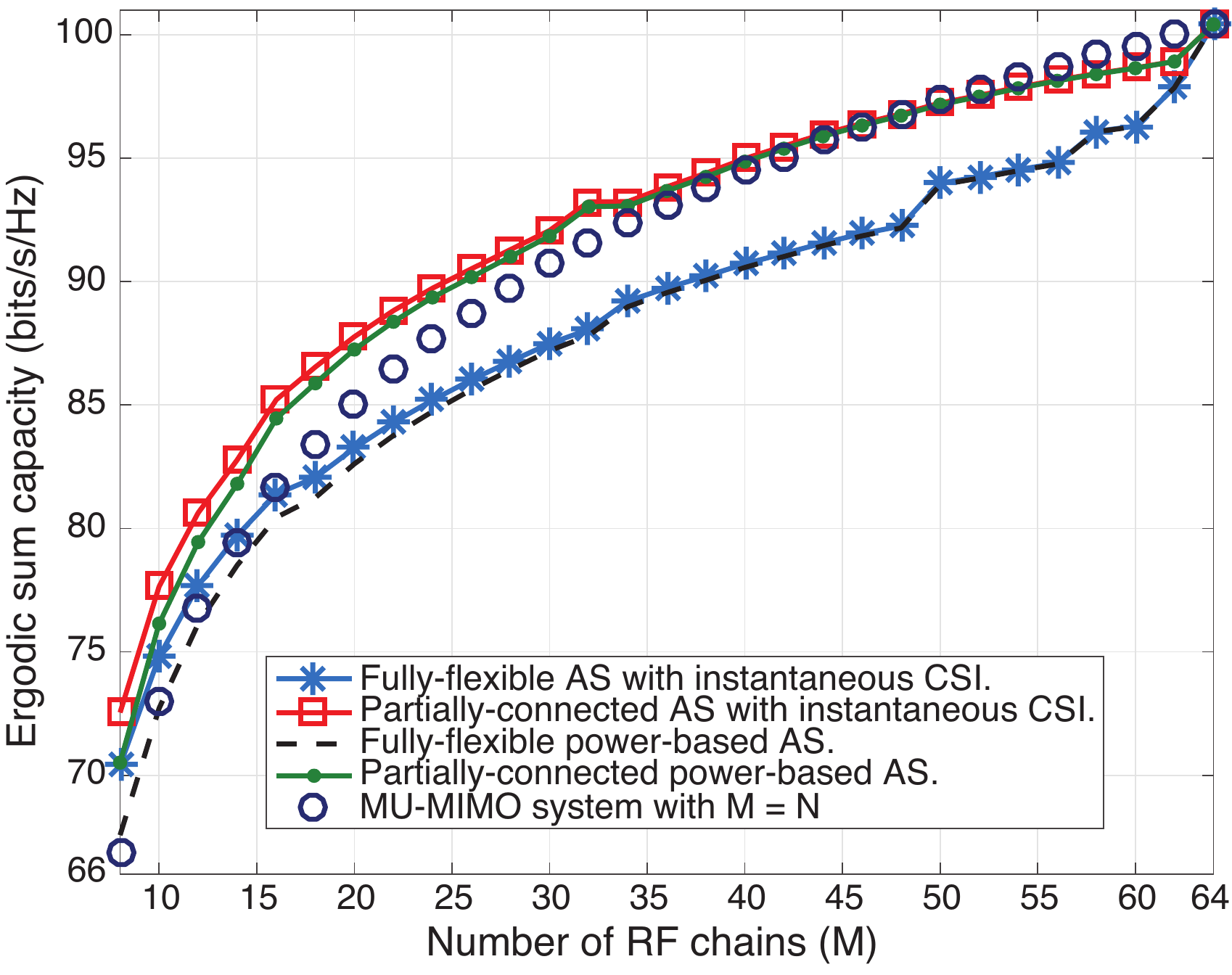}
		\caption{Ergodic sum capacity (bits/s/Hz) vs.\ $M$ for DPC with switching insertion losses. $N = 64$ for AS systems and $N = M$ for MU-MIMO without AS, $K = 8$, $\eta_{\rm dl} / \eta_{\rm coh} = 1$ and $\rho = 20$ dB. Insertion losses are considered and CSI acquisition overheads are not included.}
\label{fig:SEvsNRF_insertionLossesConsidered}
\end{figure}

\begin{figure}[!t]
	\centering
		\includegraphics[width=0.445\textwidth]{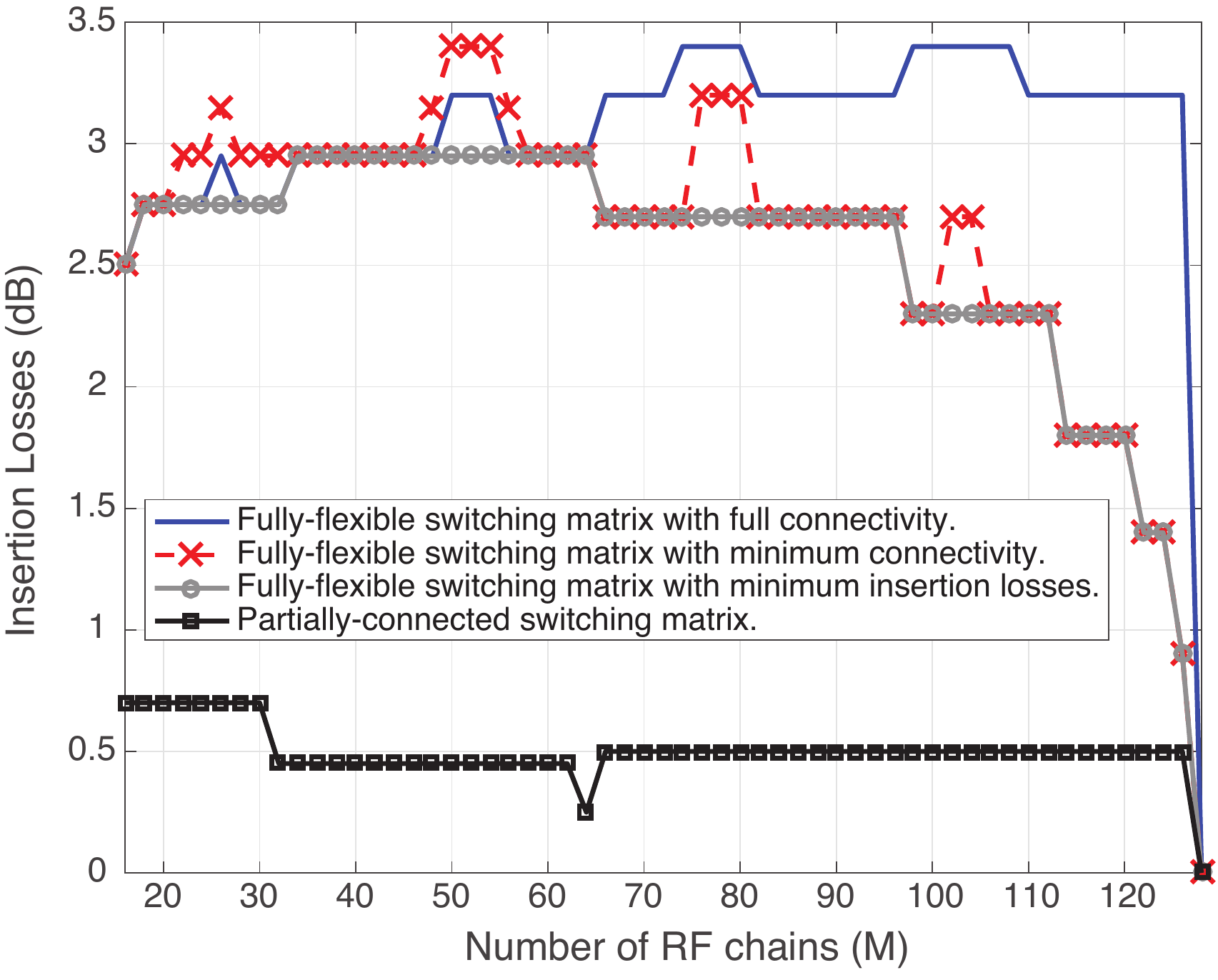}
		\caption{Insertion losses (dB) introduced by different switching architectures vs.\ $M$. $N = 128$ and basic switching losses given in Table \ref{tab:basicSwitches}.}
\label{fig:insertionLosses}
\end{figure}

\figurename~\ref{fig:SEvsNRF_insertionLossesConsidered} characterizes the impact of the switching insertion losses on the attainable ergodic sum capacity. This entails that $\rho$ represents the signal power after the PAs, while the actual transmitted power is given by $\rho / L$, where $L$ represents the switching insertion losses determined in Sec.\ \ref{sec:FFSwitching} and Sec.\ \ref{sec:PCSwitching}. In this figure we consider $N = 64$ for AS systems and $N = M$ for MU-MIMO without AS, $K = 8$, $\rho = 20$ and ignore CSI acquisition overheads. The fully-flexible switching matrix is implemented via an architecture optimized for minimum losses as detailed in Sec.\ \ref{sec:FFSwitching}. We note that the non-smooth behaviour of this system for large $M$ is a consequence of the sudden decrease in the insertion losses of the switching network, a phenomenon that can be observed in \figurename~\ref{fig:insertionLosses}. The results of \figurename~\ref{fig:SEvsNRF_insertionLossesConsidered} demonstrate the benefits of employing partially-connected switching matrices against fully-flexible implementations. This is due to a) their significantly reduced insertion losses and, as shown in \figurename~\ref{fig:spectralEfficiencyCSICost}, b) the high probability of selecting antenna combinations with similar or equal performance to those that are optimal. We also remark that larger performance differences between systems implementing AS and full MIMO with a smaller number of antennas are expected in practical implementations, due to the noticeable average power differences perceived by different antenna ports in real propagation environments \cite{7172496}.

\begin{figure}[!t]
	\centering
		\includegraphics[width=0.46\textwidth]{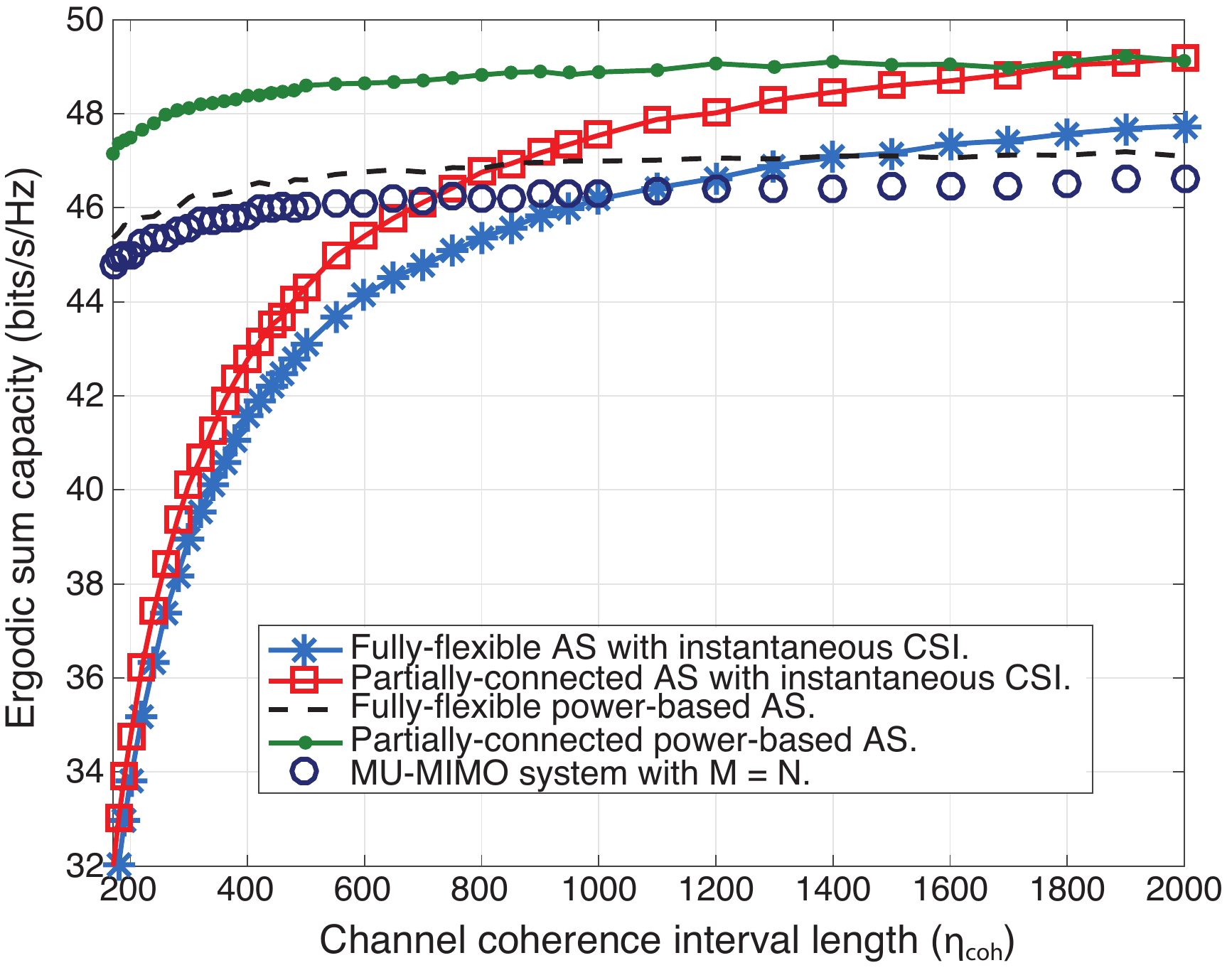}
		\caption{Ergodic sum capacity (bits/s/Hz) vs.\ $\eta_{\rm coh}$. $N = 64$ for AS systems and $N = M$ for MU-MIMO without AS, $M = K = 8$, $\rho = 20$ dB and $\eta_{\rm dl} = 0.7 \times \lceil \left( \eta_{\rm coh} - \eta_{\rm tr} \right) \rceil$. Both insertion losses and CSI acquisition overheads are considered.}
\label{fig:SEvsChannelCoherenceInterval}
\end{figure}

The results of \figurename~\ref{fig:SEvsChannelCoherenceInterval} illustrate the ergodic sum capacity against increasing values of the channel coherence interval $\eta_{\rm coh}$. We assume a system with the identical simulation parameters of \figurename~\ref{fig:SEvsNRF_insertionLossesConsidered}, $M = 8$ and $\eta_{\rm dl} = 0.7 \times \lceil \left( \eta_{\rm coh} - \eta_{\rm tr} \right) \rceil$. A maximum of $\eta_{\rm coh} = 2000$ symbols is considered, since the system behaviour is stable for larger values. This corresponds to a vehicle moving at an approximate speed of $45$ km/h at a central frequency of $2$ GHz. Overall, \figurename~\ref{fig:SEvsChannelCoherenceInterval} illustrates the transition between fast varying and slowly varying channels and allows determining the optimal selection strategy depending on $\eta_{\rm coh}$. For instance, the results of this figure demonstrate that power-based AS implemented via partially-connected switching is the solution that maximizes performance for a wide range of $\eta_{\rm coh}$ thanks to both their reduced insertion losses and shorter CSI acquisition interval. Moreover, it can be observed that acquiring accurate CSI might be advantageous for channels with large coherence intervals.

\figurename~\ref{fig:insertionLosses} represents the insertion losses of \eqref{eq:totalIL} introduced by the fully-flexible and partially-connected switching architectures considered in this paper against increasing $M$. We set $N = 128$ and consider the basic switches described in Table \ref{tab:basicSwitches} for implementing the switching network. The results of \figurename~\ref{fig:insertionLosses} clearly show the benefits of considering partially-connected architectures when compared with fully-flexible schemes, which motivates their employment when simultaneously considering their performance enhancements depicted in \figurename~\ref{fig:SEvsNRF_insertionLossesConsidered} and \figurename~\ref{fig:SEvsChannelCoherenceInterval}. For instance, the insertion losses can be reduced by up to $2.5$ dB for $M = N/2$, which is the point with minimal insertion loss for the partially-connected network and the specific case considered in \cite{7417765}. This can be explained by noting that $T_{\text{RF}} = 2$ and $T_{\text{AN}}  = 1$ for $M = N/2$. Instead, implementing a larger $M$ requires additional RF switches at the output stage of the switching network as illustrated in \figurename~\ref{fig:examplesConnectivity}(b), hence introducing additional losses in the critical signal path. However, we remark that considering a larger number of RF chains might also be required in realistic systems for satisfying specific sum rate requirements.

\figurename~\ref{fig:insertionLosses} also shows that there are non-desirable areas where the power losses can be substantially increased, if a fully-flexible architecture is preferred. Let us concentrate on understanding the behaviour these fully-flexible architectures. As expected, the fully-connected architecture designed to minimize the power losses introduces smaller insertion losses than those with different criteria. Moreover, it can be observed that there are points where minimizing the number of connections as per fully-flexible with minimum connectivity scheme also entails larger insertion losses than fully-flexible architectures with complete input-output connectivity.  This counter-intuitive behaviour arises due to the differences in the insertion loss of the basic switches and the specific number of ports $T_{\text RF}$ and $T_{\text AN}$ required at the switching matrices, as analyzed in Sec.\ \ref{sec:FFSwitching}. Overall, the results of \figurename~\ref{fig:insertionLosses} provide meaningful insights for the design of AS systems, since it can be observed that the insertion losses of different switching architectures do not follow a monotonic trend with $M$.

\begin{figure}[!t]
	\centering
		\includegraphics[width=0.473\textwidth]{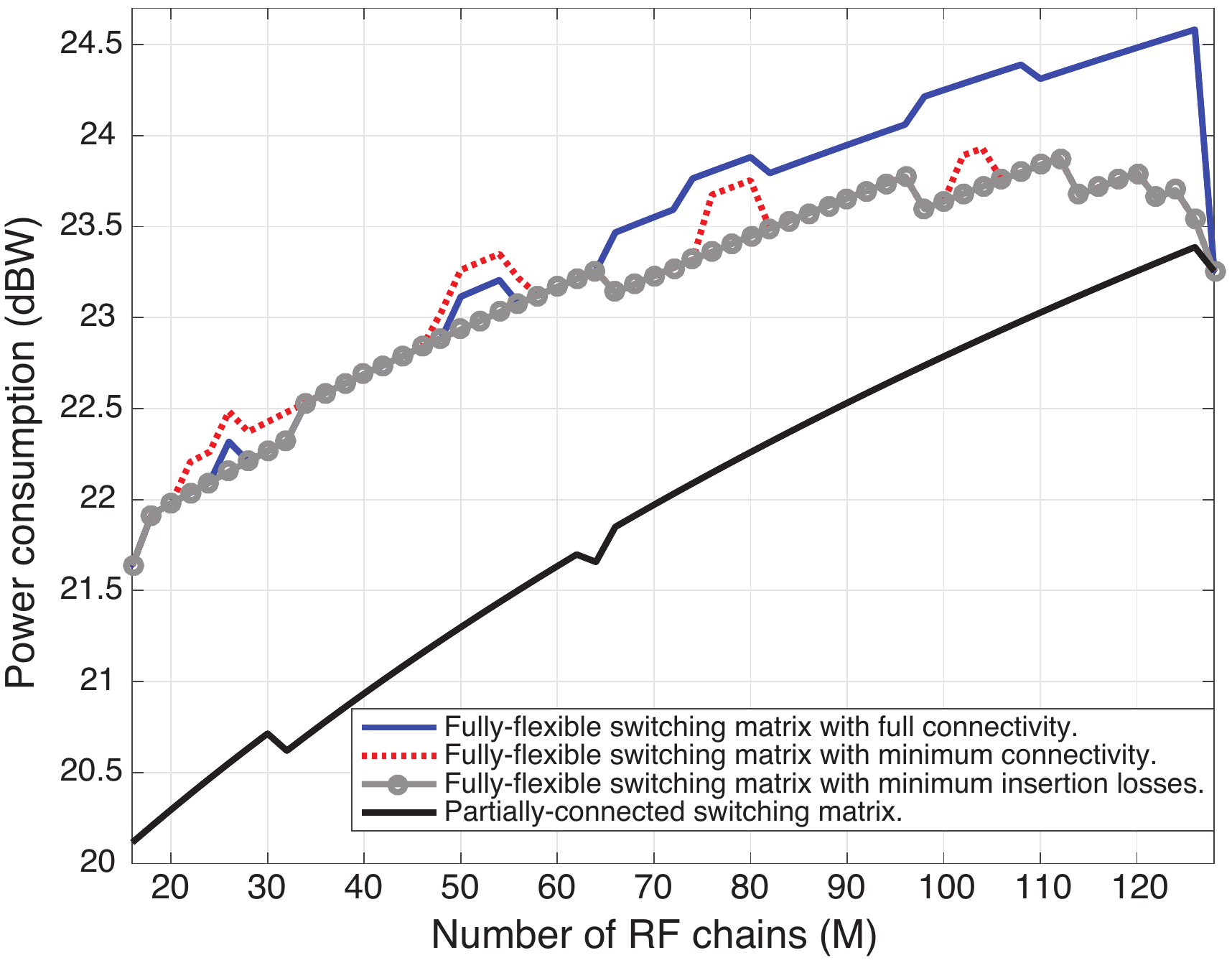}
		\caption{Total system's power consumption vs.\ $M$. $N = 128$, $\eta_{\rm tr} = K = 16$ and $\eta_{\rm coh} = 200$, $\eta_{\rm dl} = 0.7 \times \lceil \left( \eta_{\rm coh} - \eta_{\rm tr} \right) \rceil$ and $N_{\rm coh} = 1200$.}
\label{fig:totalPowerConsumption}
\end{figure}

The total system power consumption $P_{\rm tot}$ in \eqref{eq:EEFormula} is shown in \figurename~\ref{fig:totalPowerConsumption} against increasing values of $M$. Here, we have considered the values shown in Table \ref{tab:powerConsumption} and set $N = 128$, $\eta_{\rm tr} = K = 16$ and $\eta_{\rm coh} = 200$, $\eta_{\rm dl} = 0.7 \times \lceil \left( \eta_{\rm coh} - \eta_{\rm tr} \right) \rceil$. Moreover, $N_{\rm coh} = 1200$ subcarriers are assumed, which corresponds to a transmission bandwidth of $B = 20$ MHz in LTE \cite{ghosh2010fundamentals}. For simplicity, we also consider that the ADCs, DACs  only consume power and generate data to be transmitted by the DSP data interface for the fractions of time that they are active. These correspond to CSI acquisition ($\eta_{\rm tr} / \eta_{\rm coh}$) for ADCs and to downlink data transmission ($\eta_{\rm dl} / \eta_{\rm coh}$) for PAs and DACs.

The results of \figurename~\ref{fig:totalPowerConsumption} illustrate the importance of reducing the insertion losses in the switching stage, since substantial power savings over fully-flexible AS can be attained when employing the partially-connected architecture. Additionally, the results of this figure demonstrate that reducing the number of active RF chains can have substantial benefits in the overall power consumption. This is a direct consequence of the power consumed by a) the additional components required per each RF chain ($P_{\rm cir}$ and $P_{\rm DAC}, P_{\rm ADC}$), b) the additional signal processing load ($P_{\rm BB}$) and c) the increased power consumption in the data interfaces ($P_{\rm int}$) due to the increased amount of data generated \cite{erdmann2015heterogeneous}.

\begin{figure}[!t]
	\centering
		\includegraphics[width=0.45\textwidth]{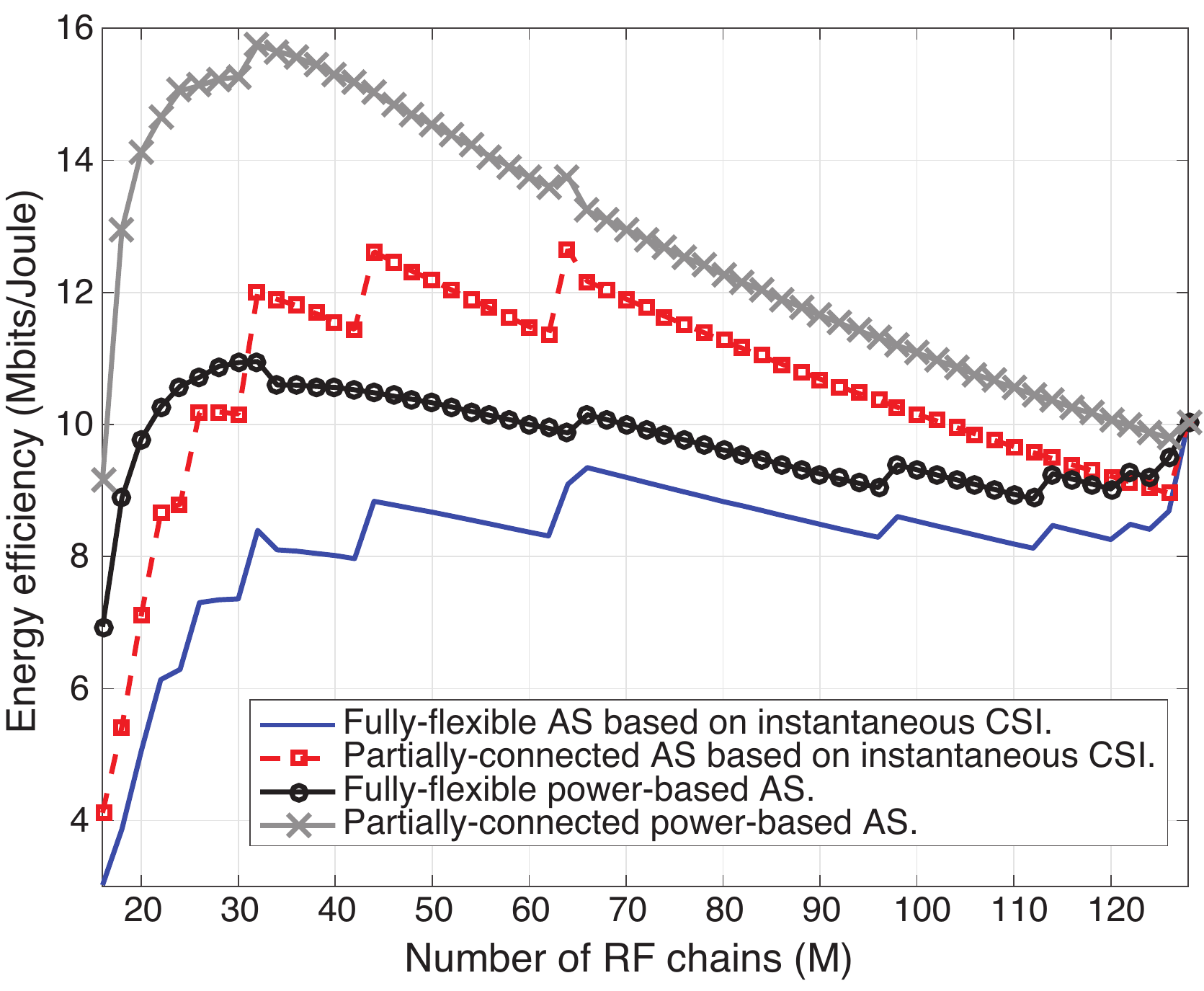}
		\caption{Energy efficiency $\xi$  vs.\ $M$ for a ZF precoding system. $N = 128$, $\eta_{\rm tr} = K = 16$ and $\eta_{\rm coh} = 200$, $\eta_{\rm dl} = 0.7 \times \lceil \left( \eta_{\rm coh} - \eta_{\rm tr} \right) \rceil$, $N_{\rm coh} = 1200$ and $\rho = 15$ dB. Insertion losses and CSI acquisition overheads are considered.}
\label{fig:energyEfficiencyZF}
\end{figure}

The energy efficiency $\xi$ in \eqref{eq:EEFormula} is shown vs.\ $M$ in \figurename~\ref{fig:energyEfficiencyZF} for a ZF precoding system with the same simulation parameters of \figurename~\ref{fig:totalPowerConsumption}. Considering standard LTE values, we assume a total system bandwidth of $B = 20$ MHz, uniform power allocation throughout the entire transmission band and, without loss of generality, we concentrate on a resource block (RB) of $B_{\rm RB} = 180$ kHz. \figurename~\ref{fig:energyEfficiencyZF} illustrates the benefits of employing both a switching network with limited connectivity and relying on power estimates for performing the AS decision. Specifically, it can be observed that the energy efficiency of the power-based partially-connected AS  is maximized for $M \approx 32$, which can be explained by noting a) the significantly reduced insertion loss when compared with smaller $M$ in \figurename~\ref{fig:insertionLosses} and b) that a large portion of the maximum attainable sum rates for $M= 128$ is already obtained for $M \approx 32$ as shown in \figurename~\ref{fig:spectralEfficiencyCSICost}. Overall, \figurename~\ref{fig:energyEfficiencyZF} demonstrates the importance of considering the insertion losses in the switching stage for energy-efficient system design.

\section{Conclusion}
\label{sec:conclusion}

In this paper, we analyze the impact of implementing different RF switching architectures in AS systems. The hardware features of a number of conventional fully-flexible switching designs are characterized and we show that they incur in significant insertion losses. For this reason we analyze switching matrices with partial connectivity, which are capable of significantly reducing the insertion losses of fully-flexible switching matrices. In this work we also provide analytical expressions to determine the performance loss introduced by designs with partial connectivity due to the limitation in the number of antenna subsets that can be selected. Overall, our numerical results show that partially-connected switching architectures a) are capable of providing significant improvements in the system's sum rates when insertion losses are considered and b) their combination with power-based antenna selection generally constitutes the most effective solution when CSI acquisition overheads are considered. Future work includes studying the impact of relying on the uplink data for performing power-based AS, incorporating hardware non-idealities in the parallel power meter chains and the implementation of partially-connected switching matrices in beam switching architectures.

\section*{Appendix A. Computation of the Joint Probabilities in \eqref{eq:jointProbabilityFormula}}

In this Appendix we outline the procedure for computing the probabilities $P \left( T_{j} \right)$ in \eqref{eq:jointProbabilityFormula} for completeness. In particular, we do this for the partially-connected architecture in \figurename~\ref{fig:examplesConnectivity}(a) for reasons of illustration, where a scheme with $M = 2$ and $N = 5$ is considered. As detailed in Sec.\ \ref{sec:performanceAnalysis}, the sets containing the combinations of the ordered column norms with non-zero probability of being selected are $\mathcal{B}^{1} = \left\{ 1, 2 \right\}$, $\mathcal{B}^{2} = \left\{ 1, 3 \right\}$ and $\mathcal{B}^{3} = \left\{ 1, 4 \right\}$. Let us start by computing $P \left( T_{1} \right)$, which is given by
\begin{align}
P \left( T_{1} \right) & = P \left( \text{selecting } \mathcal{B}^{1}_{1} \bigcap \text{selecting } \mathcal{B}^{1}_{2} \right)  = \\
& \overset{(a)}{=} P \left( \text{selecting } \mathcal{B}^{1}_{1} \right)  P \left( \text{selecting } \mathcal{B}^{1}_{2}  \vert \text{selected } \mathcal{B}^{1}_{1} \right) \\
& \overset{(b)}{=} P \left( \mathcal{B}^{1}_{1} = \left\{ 2,4 \right\} \right)  P \left( \text{selecting } \mathcal{B}^{1}_{2}  \vert \mathcal{B}^{1}_{1} = \left\{ 2,4 \right\} \right) \\
& +  P \left( \mathcal{B}^{1}_{1} = \left\{ 1, 3, 5 \right\} \right)  P \left( \text{selecting } \mathcal{B}^{1}_{2}  \vert \mathcal{B}^{1}_{1} = \left\{ 1, 3, 5 \right\} \right)  \\
& \overset{(c)}{=} \frac{2}{5} \times \frac{3}{4} + \frac{3}{5} \times \frac{2}{4} = \frac{3}{5}.
\end{align}
In the above expressions, $\overset{(a)}{=}$ is a direct application of \eqref{eq:jointProbabilityFormula}, $\overset{(b)}{=}$ divides the conditional probability of selecting the second antenna with the largest channel norm depending on whether the antenna with the largest norm were antennas $\mathcal{B}^{1}_{1} = \left\{ 1,3,5 \right\}$ or antennas $\mathcal{B}^{1}_{1} = \left\{ 2 ,4 \right\}$ and $\overset{(c)}{=}$ considers that the probability of finding $B_{i:N}, i \in \left\{ 1, \ldots, N\right\}$ at a given antenna port is equal for all antenna elements $\mathcal{N}$. 

A similar procedure can be employed to compute $P ( T_{3} )$, which reads as
\begin{align}
P \left( T_{3} \right) & = P \left( \text{selecting } \mathcal{B}^{3}_{1} \bigcap \text{selecting } \mathcal{B}^{3}_{2} \right)  = \\
& \overset{(a)}{=} P \left( \mathcal{B}^{3}_{1} = \left\{ 1, 3, 5 \right\} \right)  P \left( \text{selecting } \mathcal{B}^{3}_{2}  \vert \mathcal{B}^{3}_{1} = \left\{ 1, 3, 5 \right\} \right)  \\
& \overset{(b)}{=} \frac{3}{5}  P \left( \text{not selecting } B_{2:N} \bigcap B_{3:N} \vert \mathcal{B}^{2}_{1} = \left\{ 1, 3, 5 \right\} \right) \\
& = \frac{3}{5} \times \frac{2}{4} \times \frac{1}{3} = \frac{1}{10},
\end{align}
where $\overset{(a)}{=}$ and $\overset{(b)}{=}$ hold because the only possibility of selecting the antenna with the $\mathcal{B}^{3}_{2} = 4$-th largest norm is that the channels with the first, second and third largest norms are in antennas $\left\{ 1, 3, 5 \right\}$. Finally, $P \left( T_{2} \right) = 1 - P \left( T_{1} \right) - P \left( T_{3} \right) = \frac{3}{10}$. 

\section*{Acknowledgements}

The authors would like to thank Mr.\ Ray Kearney,  Dr.\ Pierluigi Amadori and Dr.\ Florian Pivit for their valuable comments on the paper.

\bibliographystyle{IEEEtran}
\bibliography{references}

\end{document}